\documentclass[useAMS,usenatbib]{mn2e}
\usepackage{amsmath,amssymb,verbatim,graphicx,rotating}
\usepackage{psfrag}
\usepackage{bigdelim,bigstrut,multirow}

\usepackage{color}
\newcommand{\be}{\begin{equation}}
\newcommand{\ee}{\end{equation}}
\newcommand{\bea}{\begin{eqnarray}}
\newcommand{\eea}{\end{eqnarray}}

\newcommand{\br}{\ensuremath{\mathbf{r}}}

\title[Observational rotation curves and density profiles vs. the Thomas-Fermi galaxy theory]{Observational rotation curves and density profiles vs. the Thomas-Fermi galaxy structure theory}
\author[H. J. de Vega, P. Salucci, N. G. Sanchez]{H. J. de Vega$^{1,2}$\thanks{devega@lpthe.jussieu.fr}, P. Salucci$^{3}$\thanks{salucci@sissa.it}, N. G. Sanchez$^{2}$\thanks{Norma.Sanchez@obspm.fr}\\
$^{1}$Sorbonne Universit\'es, UPMC (Univ. Paris VI), CNRS, 
Laboratoire Associ\'e au CNRS UMR 7589, \\
Tour 13-14, 4\`eme. et 5\`eme. \'etage, 
Bo\^{\i}te 126, 4, Place Jussieu, 75252 Paris, France. \\
$^{2}$Observatoire de Paris, LERMA. Laboratoire Associ\'e au CNRS UMR 8112.
 \\61, Avenue de l'Observatoire, 75014 Paris, France.\\
$^{3}$SISSA/ISAS and INFN, Trieste, Iniziativa Specifica QSKY,
via Bonomea 265, I-34136, Trieste, Italia.\\}
\begin{document}
\date{\today}
\pagerange{\pageref{firstpage}--\pageref{lastpage}} \pubyear{0000}

\maketitle

\label{firstpage}

\begin{abstract}
The Thomas-Fermi approach to galaxy structure determines selfconsistently 
the gravitational potential of the fermionic warm dark matter (WDM) given 
its distribution function $ f(E) $. This framework is appropriate for macroscopic 
quantum systems as neutron stars, white dwarfs and WDM galaxies.
Compact dwarf galaxies are near the quantum degenerate regime,
while large galaxies are in the classical Boltzmann regime.
We derive analytic {\bf scaling relations} for the main galaxy magnitudes:
halo radius $ r_h $, mass $ M_h $ and phase-space density which 
are all well reproduced by the observational data for a large variety of galaxies.
Small deviations from the exact scaling show up for compact dwarfs due to 
quantum macroscopic effects. We contrast the theoretical curves for the circular galaxy
velocities $ v_c(r) $ and density profiles $ \rho(r) $ with those
obtained from observations using the empirical Burkert profile.
Results are {\bf independent} of any WDM particle physics model,
they only follow from the gravitational interaction of the WDM particles and 
their fermionic nature. The theoretical rotation curves and density profiles 
reproduce very well the observational curves for $ r \lesssim r_h $ 
obtained from ten different and independent sets of data for galaxy masses 
from $ 5 \times 10^9 \; M_{\odot} $ till $ 5 \times 10^{11} \; M_{\odot} $.
Our normalized theoretical circular velocities and normalized density profiles 
turn to be {\bf universal} functions of $ r/r_h $ for {\bf all} galaxies. 
In addition, they agree extremely well with the observational curves described 
by the Burkert profile for $ r \lesssim 2 \; r_h $. These results show that the 
Thomas-Fermi approach correctly describes the galaxy structures.
\end{abstract}


\section{INTRODUCTION}

Dark matter (DM) is the main component of galaxies: the fraction of DM over the total
galaxy mass goes from 95\% for large dilute galaxies \citep{memo,oh,pers}
till 99.99\% for dwarf compact galaxies \citep{brodie,martin,mwal,willm,woo}.
Therefore, the study of galaxy properties is an excellent way to 
disentangle the nature of DM.

\medskip

Warm Dark Matter (WDM), that is dark matter formed by particles with masses in the keV scale
receives increasing attention today (\cite{highp,highm,cosmf} and references therein).

{\vskip 0.1cm} 

At intermediate scales $ \sim 100 $ kpc, WDM gives the {\bf correct abundance} of substructures 
and therefore WDM solves the CDM overabundance of structures at small scales.
\citep{colin,dolgov,theuns,tikho,zav,pap,lov12,lovl,ander}.
For scales larger than $ ~ 100 $ kpc, WDM yields the same results than CDM.
Hence,  WDM agrees with all the observations:
small scale as well as large scale structure observations and CMB anisotropy observations.

{\vskip 0.1cm} 

WDM simulations (as \cite{avi,colin8,vinas,theuns,lov12,lovl,sinz,zav} and many others)
are purely classical (i.e. WDM quantum dynamics is not used in those
simulations). The dynamics of DM in the simulations is worked out
classically from the classical Newton's equations or self-gravitating
hydrodynamics (also classical, without WDM quantum effects).
Quantum effects, as the DM quantum pressure, are absent is such frameworks.
Inside galaxy cores, below  $ \sim 100$ pc, WDM $N$-body classical physics simulations 
do not provide the correct structures because quantum effects are important in WDM at these scales.
Classical physics $N$-body WDM simulations without the WDM quantum pressure
exhibit cusps or small 
cores with sizes smaller than the observed cores \citep{avi,colin8,vinas,sinz}.
The relevant WDM quantum effect, as discussed in our previous articles
(Destri, de Vega, Sanchez 2013a, in what follows \citep{newas}, Destri, de Vega, Sanchez 2013b, 
in what follows \cite{astro}), is the fermionic quantum pressure. 

In WDM simulations the fact that the DM is warm appears in the 
primordial power spectrum  which is suppresed at
small scales below the free streaming length. This can be implemented
by a simple cutoff or through precise formulas fitting the WDM primordial
power spectrum (see \citep{fluct} for a recent article), and through the 
non-zero particle  velocity dispersion.
In addition, in fermionic WDM, the phase space density is bounded from above by
the Pauli principle \citep{newas}, \citep{treg}. This is the only quantum (fermionic) 
aspect of WDM implemented in
the WDM simulations. Such bound is clearly not enough to account for 
the quantum pressure of the WDM fermions, because the quantum pressure requires of a 
combination of both the Pauli principle and the Heisenberg 
principle. And therefore, the power spectrum cutoff and  the  Pauli  
bound on the phase-space density
are not enough in the simulations to enlarge the size of  the 
WDM halos  against  the gravitation attraction. That is the reason 
why in WDM simulations the core size problem persists. The presence 
of  a repulsive quantum fermionic pressure is crucial to enlarge
enough the halos against gravitation to account for the macroscopic 
core sizes.
The cutoff in the primordial power spectrum  in the WDM simulations 
is enough to account for the right number of substructures
and solve the  CDM overabundance problem.

Besides of the cutoff in the primordial power spectrum and the 
Pauli bound in the phase-space density implemented in the WDM simulations,
the crucial point is that in WDM simulations the N-body 
self-gravitating classical
evolution follows the classical Newton's equations. This dynamics in 
the inner dense regions is far away from the quantum
evolution according to the N-body Schr\"odinger equation and  does 
not contain the quantum fermionic pressure.
Instead, the Thomas-Fermi
approach corresponds to the Schr\"odinger equation in the large N 
regime and contains from the start the quantum pressure.
This is the reason why the quantum pressure is naturally
contained in the Thomas-Fermi approach and it is not included so far in the 
classical N-body WDM simulations.

The quantum pressure is well captured in the Thomas-Fermi approach \citep{newas,astro}.
The lack of quantum pressure in the WDM simulations explains why they
exhibit cusps or small cores with sizes smaller than the observed cores \citep{sinz}.
WDM predicts correct structures and cores with the right sizes for small scales 
(below the kpc scale) when its {\bf quantum} nature is taken into account \citep{newas,astro}.

{\vskip 0.1cm}

The quantum effects of matter in the inner halo regions arise because
of the quantum fermionic nature and of the quantum uncertainty principle,
the combined action of both translates into a non-zero quantum
pressure. For macroscopic systems with a large number of particles
as galaxies, this translates into macroscopic quantum effects. (Other
examples in nature are He$^3$, white dwarf stars and neutron stars).
The quantum pressure goes as the mass of the particle to the power -8/3
and it is therefore much smaller for baryons than for WDM \citep{newas}.

{\vskip 0.1cm}

The Thomas-Fermi DdVS approach applies irrespective of the WDM particle physics model.
The lower bound for the WDM particle mass $ m > 1.91 $ keV is
derived in the Thomas-Fermi approach \citep{astro} from the lightest 
known dwarf galaxies. This value is independent of the WDM particle 
physics model.

{\vskip 0.1cm}

The main fermionic WDM particle candidate is a sterile neutrino
in the keV scale. Many models of sterile neutrinos are available by now [see 
for a recent review \cite{merle}].
Another fermionic WDM particle candidate is a gravitino in the keV scale.

{\vskip 0.1cm} 

In summary, all the small structure formation constraints on the WDM 
particle mass \citep{lov12,lovl,menci,nier,pacu,pap,zav}, as well as the 
bounds from sterile neutrino decay into X-rays \citep{wat} favours 
a WDM particle mass approximately in the 2 - 3 keV range.

\medskip

Bounds on the WDM particle mass from Lyman-$\alpha$ forest data \citep{viel}
may be not so reliable since they are affected by the difficult-to-characterize 
non-linear growth of baryonic and DM structures \citep{wat}. 
Besides these systematic effects from the Lyman-$\alpha$ data, 
there are uncertainities in the WDM simulations themselves mainly 
originated from the uncertainty on the chosen 
initial velocity dispersion for the particles in the simulations whose 
effective mass is about $ 10^5 \, M_\odot = 10^{68} $ keV each, as discussed by 
several authors \citep{lov12,lovl,viel,sinz}. Namely, the effective particles 
in the WDM simulations are about $ 10^{68} $ times heavier than the real 
WDM particles. This makes diffficult to infer the initial velocity distribution 
of the effective particles from the known initial velocity distribution of the 
real WDM particles.

The Lyman-$\alpha$ mass bounds are usually given for the thermal relic mass.
This is the mass of the WDM particle if it decouples in thermal
equilibrium, which is normally not the case for sterile neutrinos.
The relation between the physical particle mass and the thermal mass has
to be worked out explicitly for each specific particle physics model.
About hundred sterile neutrino models are available today \citep{merle}
for which the Lyman-$\alpha$ bounds are not known. 
Therefore, it is not possible so far to provide precise
generic Lyman-$\alpha$ bounds on the WDM sterile neutrino mass.
At present, Lyman-$\alpha$ bounds on the WDM particle mass are only available
for a few specific particle physics models. 
For discussions on small structure formation in WDM and Lyman-$\alpha$
bounds on the WDM particle mass see \cite{highp,highm}.

\medskip

The results presented in this paper do not depend on the precise value 
of the WDM particle mass $ m $ but only on the fact that $ m $ is in the keV scale.

One can determine the keV scale of the DM particle mass $ m $ but
not its precise value within the keV scale just
from the core radius value of dilute galaxies, 
those with $ M_h > 10^6  M_\odot $ used in this paper.

The aim of this paper is to show that the obtained rotation curves and density profiles
in the Thomas-Fermi galaxy structure theory are in well agreement with
the galaxy data parametrized with Burkert profiles.

\medskip

We follow here the Thomas-Fermi approach to galaxy structure for self-gravitating 
fermionic WDM \citep{newas,astro}. This approach is especially appropriate to take into account quantum properties
of systems with large number of particles. That is, macroscopic quantum systems
as neutron stars and white dwarfs \citep{ll}.
In this approach, the central quantity to derive is the DM chemical potential $ \mu(\br) $,
which is the free energy per particle. For self-gravitating systems,
the potential $ \mu(\br) $ is proportional to the gravitational potential $ \phi(\br) $,
\be \label{potq}
  \mu(\br) =  \mu_0 - m \, \phi(\br) \; ,
\ee
$ \mu_0 $ being a constant, and 
obeys the {\bf self-consistent} and {\bf nonlinear} Poisson equation
\be\label{poisI}
\nabla^2 \mu(\br) = -4 \; \pi \; g \; G \; m^2 \; 
\int \frac{d^3p}{(2 \, \pi \; \hbar)^3} \; f\left(\frac{p^2}{2 \, m}-\mu(\br)\right) \; .
\ee
Here $ G $ is Newton's gravitational constant,  $ g $ is the number of internal degrees of freedom 
of the DM particle, $ p $ is the DM particle momentum and $ f(E) $ is the
energy distribution function. This is a semiclassical gravitational approach to 
determine selfconsistently the gravitational potential of the fermionic WDM given 
its distribution function $ f(E) $.

\medskip

In the Thomas-Fermi approach, DM dominated galaxies are considered in a stationary state.
This is a realistic situation for the late stages of structure formation
since the free-fall (Jeans) time $ t_{ff} $ for galaxies is much shorter than the age of galaxies.
$ t_{ff} $ is at least one or two orders of magnitude smaller than the age of the galaxy.

\medskip

We consider spherical symmetric configurations where eq.(\ref{poisI}) becomes an
ordinary nonlinear differential equation that determines self-consistently
the chemical potential $ \mu(r) $ and constitutes the Thomas--Fermi approach \citep{newas,astro}.
We choose for the energy distribution function a Fermi--Dirac distribution
$$
f(E) = \frac1{e^{E/E_0} + 1} \; ,
$$
where $ E_0 $ is the characteristic one--particle energy scale. $ E_0 $ plays
the role of an effective temperature scale and
depends on the galaxy mass. The Fermi--Dirac distribution function
is justified in the inner regions of the galaxy,
inside the halo radius where we find that the Thomas--Fermi density
profiles perfectly agree with the observational data modelized with the 
empirical Burkert profile.

\medskip

Observations show that the DM angular momentum is small.
In spirals we have a direct proof of this fact from their bottom
up general scenario of formation. In these objects we can
compute from observations the disk angular momentum, if
the angular momentum per unit mass is
conserved during the process of disk formation, the values
found imply that DM halos are not dominated by rotation \citep{toni}.
Therefore, the spherical symmetric approximation makes sense.
Indeed, our results confirm the consistency of such assumption.

{\vskip 0.1cm} 

In this paper spherical symmetry is considered for simplicity to
determine the essential physical galaxy properties as the classical 
or quantum nature of galaxies, compact or dilute galaxies,
the phase space density values, the
cored nature of the mass density profiles, the galaxy masses and
sizes.  It is clear that DM halos are not perfectly
spherical but describing them as spherically symmetric is a first
approximation to which other effects can be added.
In \cite{newas} we estimated the angular momentum
effect and this yields small corrections.

{\vskip 0.1cm} 

Our spherically symmetric treatment captures the essential features
of the gravitational dynamics and agree with the observations.
Notice that we are treating the DM particles quantum mechanically through
the Thomas-Fermi approach, so that expectation values are independent
of the angles (spherical symmetry) but the particles move and fluctuate
in all directions in totally non-spherically symmetric ways. 
Namely, this is more than treating purely
classical orbits for particles in which only radial motion is present.
The Thomas-Fermi approach to galaxies can be generalized to
describe non-spherically symmetric and non-isotropic situations,
by considering  distribution functions which include other
particle parameters like the angular momentum.

\medskip

The solutions of the Thomas--Fermi equations (\ref{poisI})
are characterized by the value of the chemical potential at the origin  $ \mu(0) $.
Large positive values of $ \mu(0) $ correspond to dwarf 
compact galaxies (fermions near the quantum degenerate limit),
while large negative values of $ \mu(0) $ yield large and dilute galaxies (classical 
Boltzmann regime).

{\vskip 0.1cm} 

Approaching the classical diluted limit yields larger and larger halo radii, galaxy masses
and velocity dispersions. On the contrary, in the quantum degenerate limit
we get solutions of the Thomas--Fermi equations corresponding to the {\bf minimal} halo radii, galaxy masses
and velocity dispersions. 

\medskip

The surface density 
\be\label{densuI}
\Sigma_0 \equiv  r_h  \; \rho_0  \simeq 120 \; M_\odot /{\rm pc}^2 \quad 
{\rm up ~ to } \; 10\% - 20\% \; ,
\ee
has the remarkable property of being nearly {\bf constant} and independent of 
luminosity in different galactic systems (spirals, dwarf irregular and 
spheroidals, elliptics) spanning over $14$ magnitudes in luminosity and over different 
Hubble types \citep{dona,span}. It is therefore a useful characteristic scale
to express galaxy magnitudes.

{\vskip 0.1cm} 

To reproduce the smaller observed structures the WDM particle mass should be
in the keV scale. We choose the value 2 keV as references scale
to express physical magnitudes.

\medskip

In this paper, we compute the circular velocity in the Thomas-Fermi approach using its expression
in terms of the chemical potential:
$$ 
v_c(r)  = \sqrt{\frac{G \; M(r)}{r}}  = \sqrt{- \frac{r}{m} \; \frac{d\mu}{dr}} \; ,
$$ 
On the other hand, the circular velocities of galaxies are known with precision from the observational data from
the kinematics of thousands disk galaxies and from the information arising
from other tracers of the gravitational field of galaxies as 
the dispersion velocities of spheroidals and weak lensing measurements
(\citet{sal07} and references therein). All this evidence shows that
an empirical Burkert profile 
\be\label{burI}
\rho_{B}(r) = \frac{\rho_0}{\left(1+ \displaystyle \frac{r}{r_h} \right) \; 
\left[1 +  \displaystyle \left(\frac{r}{r_h}\right)^{\! 2}\right]} \; ,
\ee
correctly reproduces the observations out to the galaxy virial radius.

{\vskip 0.1cm} 

In this paper, we contrast the observational curves for the circular 
velocities of galaxies $ V_{URC, \, h}(r) $ and the density profiles
obtained from observations using the empirical Burkert profile eq.(\ref{burI}) with
the theoretical results $ v_c(r) $ and $ \rho(r) $
arising from the resolution of the Thomas-Fermi equations.

{\vskip 0.1cm} 

Our theoretical results follow solving
the self-consistent and nonlinear Poisson equation 
eq.(\ref{poisI}) which is {\bf solely} derived 
from the purely {\bf gravitational} interaction
of the WDM particles and their {\bf fermionic} nature.
%All results are totally independent of the WDM particle physics model.
All results are valid for self-gravitating fermionic WDM particles which
are assumed stable (or with a lifetime of the order or
longuer than the Hubble time).
The non-gravitational interactions of the WDM particles are assumed weak
enough to satisfy the particle accelerator bounds and beta decay 
bounds. Except for these general
WDM particle properties, the framework described here does not 
require any particular particle physics model
of WDM production. All  the results reported here
are independent of the details of the WDM particle physics model as the
symmetry group and the values of the weak enough particle couplings.

\medskip  

The theoretical rotation curves and density profiles 
well reproduce inside the halo radius the observational curves
described by the empirical Burkert profile,
obtained from ten different and independent sets of data
for galaxy masses from $ 5 \times 10^9 \; M_{\odot} 
$ till $ 5 \times 10^{11} \; M_{\odot} $.

{\vskip 0.1cm} 

Our theoretical circular velocities and density profiles exhibit the {\bf universal}
property as the observational curves do, and in addition,
they {\bf coincide} with the observational curves 
described by the empirical Burkert profile
for $ r \lesssim 2 \; r_h $.

\medskip 

In summary, the results presented in this paper show the ability of the Thomas-Fermi approach
to correctly describe the galaxy structures.

\medskip

This paper is organized as follows. In Section 2 we present the Thomas-Fermi approach
to galaxy structure and we express the main galaxy magnitudes in terms of the solution
of the Thomas-Fermi equation and the value of the surface density $ \Sigma_0 $.
We discuss the theoretical circular velocity curves, the theoretical density profiles
and the remarkable universality of them.
In Section 3 we present the contrast between the observational and theoretical curves 
for the galaxy circular velocities and the density profiles and we discuss the 
universal property of these profiles. Section 4 is devoted to our conclusions.

\section{Galaxy properties in the Thomas-Fermi WDM approach}\label{fortoy}

We consider DM dominated galaxies in their late stages of structure formation when 
they are relaxing to a stationary situation, at least 
not too far from the galaxy center.

{\vskip 0.1cm} 

This is a realistic situation since the free-fall (Jeans) time $ t_{ff} $ for galaxies
is much shorter than the age of galaxies:
$$
t_{ff} = \frac1{\sqrt{G \; \rho_0}} = 1.49 \; 10^7 \; 
\sqrt{\frac{M_\odot}{\rho_0 \; {\rm pc}^3}} \; {\rm yr} \; .
$$
The observed central densities of galaxies yield free-fall times in the range
from 15 million years for ultracompact galaxies till 330 million years for
large dilute spiral galaxies. These free-fall (or collapse) times are small compared
with the age of galaxies running in billions of years.

{\vskip 0.1cm} 

Hence, we can consider the DM described by a time independent and non--relativistic 
energy distribution function $ f(E) $, where $ E = p^2/(2m) - \mu $
is the single--particle energy, $ m $ is the mass of the DM particle,
$ \mu $ is the chemical potential \citep{newas,astro},
related to the gravitational potential $ \phi(\br) $ by eq.(\ref{potq}).

{\vskip 0.1cm} 

In the Thomas--Fermi approach, $ \rho(\br) $ is expressed as a function of $\mu(\br)$ through the
standard integral of the DM phase--space distribution function over the momentum
\be \label{den}
  \rho(\br) = \frac{g \, m}{2 \, \pi^2 \, \hbar^3} \int_0^{\infty} dp\;p^2 
  \; f\left(\displaystyle \frac{p^2}{2m}-\mu(\br)\right)\; , 
\ee
where $ g $ is the number of internal degrees of freedom of the DM particle,
with $ g = 1 $ for Majorana fermions and $ g = 2 $ for Dirac fermions. For
definiteness, we will take $g=2$ in the sequel.

\medskip

We will consider spherical symmetric configurations. Then, 
%In the spherical symmetric case 
the Poisson equation for $ \phi(r) $ takes the self-consistent form
\bea \label{pois}
&&\frac{d^2 \mu}{dr^2} + \frac2{r} \; \frac{d \mu}{dr} = - 4\pi \, G \, m \, \rho(r) =
\cr \cr
&& = - \frac{4 \; G \; m^2}{\pi \; \hbar^3} \int_0^{\infty} dp\;p^2 
  \; f\left(\displaystyle \frac{p^2}{2m}-\mu(r)\right)\; , 
\eea
where $ G $ is Newton's constant and $ \rho(r) $ is the DM mass density. 

\medskip

Eq. (\ref{pois}) provides an ordinary {\bf nonlinear}
differential equation that determines {\bf self-consistently} the chemical potential $ \mu(r) $ and
constitutes the Thomas--Fermi approach \citep{newas,astro}. This is a semi-classical approach
to galaxy structure in which the quantum nature of the DM particles is taken into account through
the quantum statistical distribution function $ f(E) $.

\medskip

The DM pressure and the velocity dispersion can also be expressed as 
integrals over the DM phase--space distribution function as
\bea \label{P}
&&  P(r) = \frac{1}{3 \, \pi^2 \,m\,\hbar^3} \int_0^{\infty} dp\;p^4 
  \,f\left(\displaystyle \frac{p^2}{2m}-\mu(r)\right) \;  , \\  \cr  \cr 
&& <v^2>(r) = \frac1{m^2} \frac{\int_0^{\infty} dp\;p^4 
  \,f\left(\displaystyle \frac{p^2}{2m}-\mu(r)\right)}{\int_0^{\infty} dp\;p^2 
  \; f\left(\displaystyle \frac{p^2}{2m}-\mu(r)\right)} = 3 \; \frac{P(r)}{\rho(r)}
\; . \nonumber
\eea
The fermionic DM mass density $ \rho $ is bounded at the origin 
due to the Pauli principle \citep{newas} which implies 
the bounded boundary condition at the origin
\be\label{ori}
  \frac{d \mu}{dr}(0) = 0 \; .
\ee
We see that $\mu(r)$ fully characterizes the DM halo structure in this
Thomas--Fermi framework. The chemical potential is monotonically decreasing in $ r $ 
since eq.(\ref{pois}) implies
\be\label{dmu}
\frac{d\mu}{dr} = -\frac{G\,m\,M(r)}{r^2} \quad,\qquad  
  M(r) = 4\pi \int_0^r dr'\, r'^2 \, \rho(r') \; .
\ee
In this semi-classical framework the stationary energy distribution function $
f(E) $ must be given. We consider the Fermi--Dirac distribution 
\be\label{FD}
  f(E) = \Psi_{\rm FD}(E/E_0) = \frac1{e^{E/E_0} + 1} \; ,
\ee
where the characteristic one--particle energy scale $ E_0 $ in the DM halo
plays the role of an effective temperature. The value of $ E_0 $ depends on the galaxy mass.
In neutron stars, where the neutron mass is about six orders of magnitude larger
than the WDM particle mass, the temperature can be approximated by zero.
In galaxies, $ E_0 \sim m \; <v^2> $ turns to be non-zero but small in the range: 
$ 10^{-3} \; {\rm K} \lesssim E_0  \lesssim 50 $ K which reproduce the observed velocity
dispersions for $ m \sim 2 $ keV. The smaller values of $ E_0 $ correspond to compact
dwarf galaxies and the larger values of $ E_0 $ are for large and dilute galaxies.

{\vskip 0.1cm} 

Notice that for the relevant galaxy physical magnitudes, 
the Fermi--Dirac distribution function gives similar results 
to out of equilibrium distribution functions \citep{newas}.

{\vskip 0.1cm} 

The choice of $ \Psi_{\rm FD} $ is justified in the inner regions,
where relaxation to thermal equilibrium is possible. Far from the  origin
however, the Fermi--Dirac distribution as its classical counterpart, the isothermal sphere,
produces a mass density tail $ 1/r^2 $ that overestimates the observed tails of the 
galaxy mass densities. Indeed, the
classical regime $ \mu/E_0 \to -\infty $ is attained for large distances $ r $
since eq.(\ref{dmu}) indicates that $ \mu(r) $ is always monotonically decreasing with $ r $.

{\vskip 0.1cm} 

More precisely, large positive values of the chemical potential at the origin 
correspond to the degenerate fermions limit which is the extreme quantum case and oppositely, 
large negative values of the chemical potential at the origin gives the diluted case which 
is the classical regime. The quantum degenerate regime describes dwarf and compact galaxies while
the classical and diluted regime describes large and diluted galaxies.
In the classical regime, the Thomas-Fermi equation
(\ref{pois})-(\ref{ori}) become the equations for a self-gravitating Boltzmann gas.

\medskip

It is useful to introduce dimensionless variables $ \xi , \; \nu(\xi) $ 
\be\label{varsd}
 r = l_0 \; \xi \qquad , \qquad \mu(r) =  E_0 \;  \nu(\xi) \; , 
\ee
where $ l_0 $ is the characteristic length that emerges from the dynamical equation (\ref{pois}):
\bea\label{varsd2}
&& l_0 \equiv  \frac{\hbar}{\sqrt{8\,G}} 
\left[\frac{9 \, \pi \; I_2(\nu_0)}{m^8\,\rho_0}\right]^{\! \! \frac16} 
  = R_0 \; \left(\frac{2 \, {\rm keV}}{m}\right)^{\! \! \frac43}  \; 
  \left[\frac{I_2(\nu_0)}{\rho_0} \; \frac{M_\odot}{{\rm pc}^3}\right]^{\! \! \frac16} 
  \;, \cr \cr
&& R_0 = 7.425 \; \rm pc  \; ,
\eea
and
\bea\label{dfI}
&& I_n(\nu) \equiv (n+1) \; \int_0^{\infty} y^n \; dy \; \Psi_{FD}(y^2 -\nu) \quad  , \cr \cr
&& n = 1, 2 , \ldots\; , \quad  , \quad \nu_0 \equiv \nu(0) \quad  , \quad \rho_0 = \rho(0) \; ,
\eea
where we use the integration variable $ y \equiv p / \sqrt{2 \, m \;  E_0} $.

\medskip

Then, in dimensionless variables, the self-consistent Thomas-Fermi equation 
(\ref{pois}) for the chemical potential $ \nu(\xi) $ takes the form
\be\label{nu}
\frac{d^2 \nu}{d\xi^2} + \frac2{\xi} \; \frac{d \nu}{d\xi} = - I_2(\nu)
\quad ,  \quad \nu'(0) = 0 \quad .
\ee
We solve eq.(\ref{nu}) numerically by using as independent variable $ u \equiv \ln \xi $
and then applying the fourth-order Runge-Kutta method. We solve eq.(\ref{nu}) for a broad
range of values $ \nu_0 \equiv \nu(0) $, from negative values $ \nu_0 \lesssim -5 $ describing
galaxies in dilute regimes to positive values $ \nu_0 \gtrsim 1 $ corresponding to compact
dwarf galaxies.

\medskip

We find the main physical galaxy magnitudes, such as the
mass density $ \rho(r) $, the velocity dispersion $ \sigma^2(r) = v^2(r)/3 $ and the pressure 
$ P(r) $, which are all $r$-dependent as: 
\bea\label{gorda}
&& \rho(r) = \rho_0 \; \frac{I_2(\nu(\xi))}{I_2(\nu_0)} 
\quad , \quad  \rho_0 = \frac1{3 \; \pi^2} \; 
\frac{m^4}{\hbar^3} \left(\frac{2\; E_0}{m}\right)^{3/2}
  I_2(\nu_0) \; , \cr \cr \cr 
&& \sigma(r) = \sqrt{\frac{2 \; E_0}{5 \; m} \; \frac{I_4(\nu(\xi))}{I_2(\nu(\xi))}} =
\\ \cr \cr 
&& = 1.092 \; \sqrt{\frac{I_4(\nu(\xi))}{I_2(\nu(\xi))}} \;
\left(\frac{2 \, {\rm keV}}{m}\right)^{\! \! \frac43} \; 
\left[\frac{\rho_0}{I_2(\nu_0)}\; \frac{{\rm pc}^3}{M_\odot}\right]^{\! \! \frac13} \; 
\; \frac{\rm km}{\rm s} \;  , \\ \cr \cr 
&& P(r) =  \frac{2 \; E_0}{5\,m} \; \rho_0 \; \frac{I_4(\nu(\xi))}{I_2(\nu_0)} = 
\cr \cr \cr 
&& = \frac1{5} \; \left(\frac{3 \, \pi^2 \; \hbar^3}{m^4}\right)^{\! \! \frac23}
  \left[\frac{\rho_0}{I_2(\nu_0)}\right]^{5/3} \; I_4(\nu(\xi)) \;, 
\eea
As a consequence, from eqs.(\ref{dmu}), (\ref{varsd}), (\ref{varsd2}), (\ref{nu}) and (\ref{gorda})
the total mass $ M(r) $ enclosed in a sphere of radius $ r $ and
the phase space density $ Q(r) $ turn to be
\bea\label{cero} 
&& M(r) = 4 \, \pi \; \frac{\rho_0\; l_0^3}{I_2(\nu_0)}\,\int_0^{\xi}
    dx\, x^2 \,I_2(\nu(x)) = \cr \cr \cr 
&& = 4 \, \pi \; \frac{\rho_0 \; l_0^3}{I_2(\nu_0)} \;
    \xi^2 \; |\nu'(\xi)| = \cr \cr \cr
&&=  M_0 \; \xi^2 \; |\nu'(\xi)|
    \; \left(\frac{{\rm keV}}{m}\right)^{\! \! 4} \; 
\sqrt{\frac{\rho_0}{I_2(\nu_0)}\; \frac{{\rm pc}^3}{M_\odot}} \; , 
\cr \cr 
&& M_0 = 4 \; \pi \; M_\odot \; \left(\frac{R_0}{\rm pc}\right)^{\! \! 3} 
    = 0.8230 \; 10^5 \; M_\odot \; ,\\ \cr 
&& Q(r) \equiv \frac{\rho(r)}{\sigma^3(r)} = 3 \; \sqrt3 \; \frac{\rho(r)}{<v^2>^\frac32(r)} =
 \cr \cr
&& = \frac{\sqrt{125}}{3 \; \pi^2} \; \; \frac{m^4}{\hbar^3}  \;
\frac{I_2^{\frac{5}{2}}(\nu(\xi))}{I_4^{\frac32}(\nu(\xi))} \label{Qcero} \; .
\eea
We have systematically eliminated the energy scale $ E_0 $ 
in terms of the central density $ \rho_0 $ through eq.(\ref{gorda}). 
Notice that $ Q(r) $ turns to be independent of $ E_0 $ and therefore from $ \rho_0 $.

\medskip

%Besides the virial galaxy radius $ R_{vir} = l_0 \; \xi_{vir} $, 
We define the core size $ r_h $ of the halo by analogy with the empirical Burkert density profile as
\be\label{onequarter}
\frac{\rho(r_h)}{\rho_0} = \frac14 \quad , \quad  r_h = l_0 \; \xi_h \; .
\ee

\medskip

It must be noticed that the surface density 
 \be\label{densu}
\Sigma_0 \equiv  r_h  \; \rho_0  \; ,
\ee
is found nearly {\bf constant} and independent  of 
luminosity in  different galactic systems (spirals, dwarf irregular and 
spheroidals, elliptics) 
spanning over $14$ magnitudes in luminosity and  over different 
Hubble types. More precisely, all galaxies seem to have the same value 
for $ \Sigma_0 $, namely $ \Sigma_0 \simeq 120 \; M_\odot /{\rm pc}^2 $
up to $ 10\% - 20\% $ \citep{dona,span,kor}.   
It is remarkable that at the same time 
other important structural quantities as $ r_h , \; \rho_0 $, 
the baryon-fraction and the galaxy mass vary orders of magnitude 
from one galaxy to another.

{\vskip 0.1cm} 

The constancy of $ \Sigma_0 $ seems unlikely to be a mere coincidence and probably
reflects a physical scaling relation between the mass and halo size of galaxies.
It must be stressed that $ \Sigma_0 $ is the only dimensionful quantity
which is constant among the different galaxies.

\medskip

We use here the dimensionful quantity $ \Sigma_0 $ to set the energy scale 
in the Thomas-Fermi approach. That is, we replace the central density $ \rho_0 $
in eqs.(\ref{varsd}), (\ref{varsd2}), (\ref{gorda}) and (\ref{cero}) 
in terms of $ \Sigma_0 $ eq.(\ref{densu}) with the following results
\bea\label{E0}
&& l_0 = \frac{\hbar^\frac65}{G^\frac35} \; \left(\frac{9 \; \pi}{512}\right)^{\! \! \frac15}
\; \left[\frac{\xi_h \; I_2(\nu_0)}{m^8 \; \Sigma_0}\right]^{\! \! \frac15} =
\cr \cr\cr
&& = 4.2557 \; \left[\xi_h \; I_2(\nu_0)\right]^{\! \frac15} \;
\left(\frac{2 \, {\rm keV}}{m}\right)^{\! \! \frac85} \; 
\left(\frac{120 \; M_\odot}{\Sigma_0 \;  {\rm pc}^2}\right)^{\! \! \frac15}  \;
{\rm pc}\; , \cr \cr\cr
&& E_0 = \hbar^\frac65 \; \frac{G^\frac25}{m^\frac35} \; \left(18 \; \pi^6\right)^{\! \frac15}
\; \left[\frac{\Sigma_0}{\xi_h \; I_2(\nu_0)}\right]^{\! \! \frac45} =
\cr \cr\cr
&& = \frac{7.12757 \; 10^{-3}}{\left[\xi_h \; I_2(\nu_0)\right]^{\frac45}} \;  
\left(\frac{2 \, {\rm keV}}{m}\right)^{\! \! \frac35} \; 
\left(\frac{\Sigma_0 \;  {\rm pc}^2}{120 \; M_\odot}\right)^{\! \! \frac45}  \; {\rm K} \; ,
\eea
and
\bea\label{rMr}
\! \! \! \! r \! \! \! \!& \! \!\! = \! \!\! & \! \!\! \! 4.2557 \;  \xi \; 
\left[ \xi_h \; I_2(\nu_0) \right]^{\frac15} \;
 \left(\frac{2 \, {\rm keV}}{m}\right)^{\! \! \frac85}  \; 
\left(\frac{120 \; M_\odot}{\Sigma_0 \;  {\rm pc}^2}\right)^{\! \! \frac15} {\rm pc} \\ \cr\cr
 \rho(r) &=& 18.1967 \; \frac{I_2(\nu(\xi))}{\left[\xi_h \; I_2(\nu_0)\right]^{\! \frac65}}
\; \left(\frac{m}{2 \, {\rm keV}}\right)^{\! \! \frac85} \; 
\left(\frac{120 \; M_\odot}{\Sigma_0 \;  {\rm pc}^2}\right)^{\! \! \frac65} \;
\frac{M_\odot}{{\rm pc}^3} \; , \label{rhor}\cr \cr\cr
 \! \! \! \! M(r)\! \! \! \! &\! \! \! \!=\! \! \! \!&\! \! \! \!
 \frac{27312 \; \xi^2}{\left[ \xi_h \; I_2(\nu_0) \right]^{\frac35}}
|\nu'(\xi)|  \left(\frac{2 \, {\rm keV}}{m}\right)^{\! \! \frac{16}5} 
\left(\frac{\Sigma_0 \;  {\rm pc}^2}{120 \; M_\odot}\right)^{\! \! \frac35}  M_\odot .
\label{MR}
\eea
For a fixed value of the surface density $ \Sigma_0 $, the 
solutions of the Thomas-Fermi eqs.(\ref{nu}) are parametrized by a single
parameter: the dimensionless chemical potential at the center $ \nu_0 $.
That is, $ \nu_0 $ is determined by the value of the halo galaxy mass 
\be \label{mhache}
M_h \equiv M(r_h) 
\ee

\begin{figure}
\begin{turn}{-90}
\psfrag{"RMr0lin.dat"}{Dilute regime of Thomas-Fermi}
\psfrag{"RMrh.dat"}{Exact Thomas-Fermi}
\psfrag{Log Halo Mass}{$ \log_{10} \left[ {\hat M}_h/M_\odot \right] $}
\psfrag{Log Halo radius}{$ \log_{10}\left[ {\hat r}_h/{\rm pc}\right] $}
\includegraphics[height=9.cm,width=9.cm]{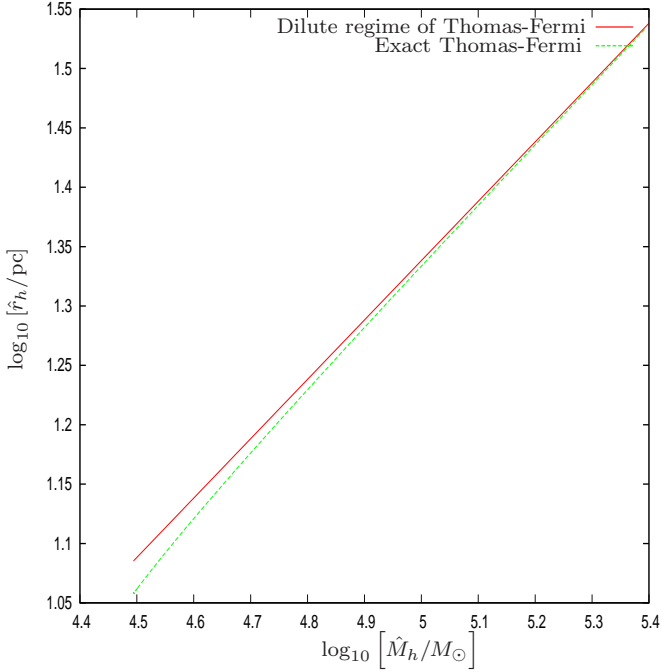}
\end{turn}
\caption{The ordinary logarithm of the theoretical halo radius $ {\hat r}_h $ 
vs. the  ordinary logarithm of the halo mass $ {\hat M}_h $ for small galaxy masses
in the Thomas-Fermi approach from eqs.(\ref{onequarter}),  (\ref{MR}),
(\ref{mhache}), (\ref{rsom}) and (\ref{Msom}) 
(dashed green line) and the dilute regime eq.(\ref{dilu})-(\ref{Mhrh}) (red continuous line).
$ {\hat r}_h $ and $ {\hat M}_h $ are defined by eqs.(\ref{rsom})-(\ref{Msom}).
The dilute regime approximates very well the exact Thomas-Fermi results,
$ r_h $ follows the square-root of $ M_h $ scaling behaviour of the classical regime 
eq.(\ref{dilu})-(\ref{Mhrh}). This is so even near the fermion degenerate quantum limit.}
\label{Rmhrh}
\end{figure}

In the classical dilute limit, $ \nu_0 \lesssim -5 $, the analytic expressions
for the main galaxies magnitudes are given by:
\bea\label{dilu}
&& \xi_h = \frac{3.147473}{e^{\nu_0/2}} \quad , \quad \left. \frac{d\nu}{d \ln \xi}\right|_{\xi_h} = -1.839957  
\quad , \\ \cr \cr 
&& M_h = 1.75572 \; \Sigma_0 \; r_h^2 \quad , \cr \cr\cr 
&&  M_h = \frac{67011.4}{e^{\frac45 \,\nu_0}} \; \left(\frac{2 \, {\rm keV}}{m}\right)^{\! \! \frac{16}5} \;
 \left(\frac{\Sigma_0 \;  {\rm pc}^2}{120 \; M_\odot}\right)^{\! \! \frac35}  \;  M_\odot
\; , \cr \cr\cr
&& r_h =  68.894 \; \sqrt{\frac{M_h}{10^6 \; M_\odot}} \;
\sqrt{\frac{120\; M_\odot}{\Sigma_0 \;  {\rm pc}^2}} \;\; \; {\rm pc} \label{Mhrh}
\; ,  \\ \cr \cr
&& r = 22.728 \; \xi \; \left(\frac{M_\odot}{M_h}\right)^{\! \frac18} 
\left(\frac{2 \, {\rm keV}}{m}\right)^2 
\left(\frac{120\; M_\odot}{\Sigma_0 \;  {\rm pc}^2}\right)^{\! \frac18} {\rm pc}\; ,  \\ \cr \cr
&& \rho(r) = 5.195045 \; \left(\frac{M_h}{10^4 \;  M_\odot}\right)^{\! \! \frac34} \;
\left(\frac{m}{2 \, {\rm keV}}\right)^4 \; 
\left(\frac{\Sigma_0 \;  {\rm pc}^2}{120 \; M_\odot}\right)^{\! \! \frac34} \times \cr \cr 
&& e^{\nu(\xi)} \; \; \frac{M_\odot}{{\rm pc}^3} \; ,\\ \cr \cr
&&  M(r) = 179.30 \; \left(\frac{M_h}{M_\odot}\right)^{\! \! \frac34} \;
\left(\frac{2 \, {\rm keV}}{m}\right)^2 \;
\left(\frac{\Sigma_0 \;  {\rm pc}^2}{120 \; M_\odot}\right)^{\! \! \frac38} \times \cr \cr 
&& \xi \; \left|\frac{d\nu(\xi)}{d \ln \xi}\right| \;  M_\odot  \; , \\ \cr \cr
&& Q(0) = 1.2319 \; \left(\frac{10^5 \;  M_\odot}{M_h}\right)^{\! \! \frac54} \;
\left(\frac{\Sigma_0 \;  {\rm pc}^2}{120 \; M_\odot}\right)^{\! \! \frac34} \; {\rm keV}^4\label{qh} \; .
\eea
These equations are accurate for $ M_h \gtrsim 10^6 \;  M_\odot $.
We see that they exhibit a {\bf scaling} behaviour for $ r_h $ vs. $ M_h $,  $ Q_h $ and
$ Q(0) $ vs. $ M_h $ and $ M_h $ vs. the fugacity at the center $ z_0 = e^{\nu_0} $. These scaling behaviours
are {\bf very accurate} except near the degenerate limit as shown by fig. \ref{Rmhrh}.

{\vskip 0.1cm} 

It must be stressed that (i) the scaling relations eqs.(\ref{dilu})-(\ref{qh}) 
are a consequence solely of the self-gravitating
interaction of the fermionic WDM and (ii) the value of the WDM particle mass $ m \simeq 2 $ keV appears
in the proportionality factors which is therefore confirmed by the galaxy data
(see figs. \ref{mhrh} and \ref{q}).

{\vskip 0.1cm} 

In eqs.(\ref{rMr})-(\ref{qh}) we use the surface density $ \Sigma_0 $ as energy scale
to express the theoretical results [we used the central density $ \rho_0 $ in 
refs. \citep{newas,astro}]. It is highly remarkable that our theoretical
results {\bf reproduce} the observed DM halo properties with {\bf good precision}.

\medskip

The limit opposite limit, $ \nu_0 \gtrsim 1 $ is the extreme quantum limit corresponding to degenerate
WDM fermions. The galaxy mass and halo radius take in the degenerate limit their {\bf minimum } values
\bea
&& r_h^{min} = 11.3794 \; \left(\frac{2 \, {\rm keV}}{m}\right)^{\! \! \frac85}  \; 
\left(\frac{120 \; M_\odot}{\Sigma_0 \;  {\rm pc}^2}\right)^{\! \! \frac15} \; {\rm pc} \; , \cr \cr
&&  M_h^{min} = 30998.7 \; \left(\frac{2 \, {\rm keV}}{m}\right)^{\! \! \frac{16}5} \;
 \left(\frac{\Sigma_0 \;  {\rm pc}^2}{120 \; M_\odot}\right)^{\! \! \frac35}  \;  M_\odot \; ,
\eea
while the phase-space density $ Q(r) $ takes its {\bf maximum} value
\be
Q_h^{max} = 16 \; \frac{\sqrt{125}}{3 \; \pi^2} \;  \left(\frac{m}{2 \, {\rm keV}}\right)^4 \; {\rm keV}^4 =
6.041628 \; \left(\frac{m}{2 \, {\rm keV}}\right)^4 \; {\rm keV}^4 \label{qhdeg} \; .
\ee
But the Thomas-Fermi equations provide a whole 
continuous range of galaxy solutions above the degenerate limit
as discussed above and in \cite{newas,astro}.

\medskip

The degenerate limit corresponds to $ E_0 = 0 $. In the classical
dilute limit $ M_h \gtrsim 10^6 \;  M_\odot , \; E_0 $ runs approximately 
from 0.02 K to 20 K.

\subsection{The galaxy circular velocities}

We consider now the circular velocity $ v_c(r) $ defined through the virial theorem as
\be\label{vci}
 v_c(r) \equiv \sqrt{\frac{G \; M(r)}{r}} \; .
\ee
The circular velocity is directly related by eq.(\ref{dmu}) 
to the derivative of the chemical potential as
$$
 v_c(r) = \sqrt{- \frac{r}{m} \; \frac{d\mu}{dr}} \; ,
$$
which in dimensionless variables takes the form
$$
v_c(r) = \sqrt{-\frac{E_0}{m} \; \frac{d\nu}{d\ln \xi}} \; .
$$
Expressing the energy scale $ E_0 $ in terms of the surface density using
eq.(\ref{E0}) we have for the circular velocity the explicit expression
\be\label{vtf}
v_c(r) = 5.2537 \; \frac{\sqrt{-\xi \; \nu'(\xi)}}{\left[ \xi_h \; I_2(\nu_0) \right]^{\frac25}}
\; \left(\frac{2 \, {\rm keV}}{m}\right)^{\! \! \frac45} \;
\left(\frac{\Sigma_0 \;  {\rm pc}^2}{120 \; M_\odot}\right)^{\! \! \frac25}  \;
\frac{\rm km}{\rm s} \; .
\ee
In the dilute Boltzmann regime the circular velocity at the core radius $ r_h $
scales as the power 1/4 of the galaxy halo mass $ M_h $:
\be\label{vcrh}
v_c(r_h) = 7.901 \; \left(\frac{M_h}{10^6 \;  M_\odot}\right)^{\! \! \frac14} \;
\left(\frac{\Sigma_0 \;  {\rm pc}^2}{120 \; M_\odot}\right)^{\! \! \frac14}  \;
\; \frac{\rm km}{\rm s} \; .
\ee

\begin{figure}
\begin{turn}{-90}
\psfrag{"5urc.dat"}{URC from Observations} 
\psfrag{"Lvcirc1.dat"}{$ M_h = 5.1 \; 10^9 \; M_{\odot} $: Theory}
\psfrag{"Lvcirc2.dat"}{$ M_h = 8.4 \; 10^9\; M_{\odot} $: Theory}
\psfrag{"Lvcirc3.dat"}{$ M_h = 1.4 \; 10^{10}\; M_{\odot} $: Theory}
\psfrag{"Lvcirc4.dat"}{$ M_h = 2.3 \; 10^{10}\; M_{\odot} $: Theory}
\psfrag{"Lvcirc5.dat"}{$ M_h = 3.8 \; 10^{10} \; M_{\odot} $: Theory}
\psfrag{"Lvcirc6.dat"}{$ M_h = 6.4 \; 10^{10} \; M_{\odot} $: Theory}
\psfrag{"Lvcirc7.dat"}{$ M_h = 1.1 \; 10^{11}\; M_{\odot} $: Theory}
\psfrag{"Lvcirc8.dat"}{$ M_h = 1.8 \; 10^{11}\; M_{\odot} $: Theory}
\psfrag{"Lvcirc9.dat"}{$ M_h = 3.0 \; 10^{11}\; M_{\odot} $: Theory}
\psfrag{"Lvcirc10.dat"}{$ M_h = 5.2 \; 10^{11} \; M_{\odot} $: Theory}
\psfrag{equis}{$ x = r/r_h $}
\psfrag{UcirN}{$ U(x) = v_c(r)/ v_c(r_h) $}
\includegraphics[height=9.cm,width=9.cm]{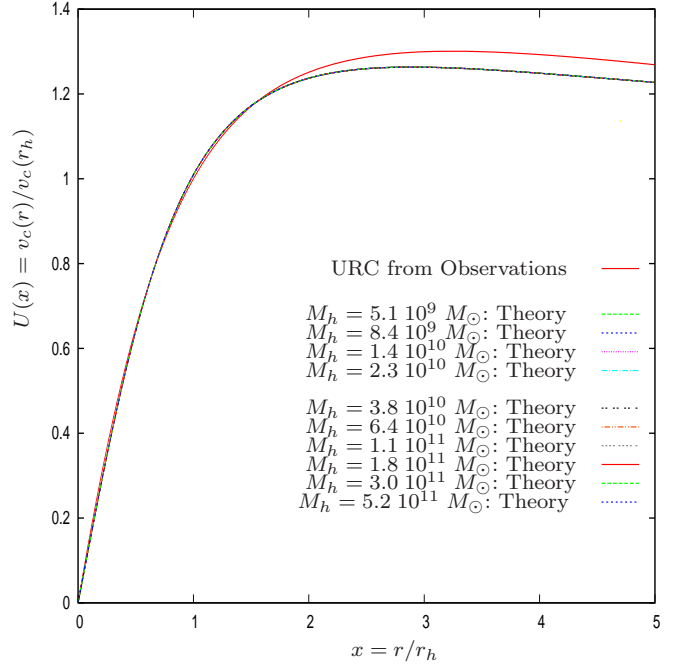}
\end{turn}
\caption{Normalized circular velocities $ U(x) = v_c(r)/ v_c(r_h) $ vs. 
$ x = r/r_h $ from the observational data eq.(\ref{ubur}) 
and from the theoretical Thomas-Fermi formula eq.(\ref{vtf}).
The theoretical curves from the Thomas-Fermi approach
for ten different galaxy masses all {\bf fall} one into each other providing 
an {\bf universal} rotation curve which practically
{\bf coincides } with the observational universal curve URC
for $ x = r / r_h \lesssim 2 $.}
\label{urc}
\end{figure}

It is important to consider the circular velocity normalized to unit
at the core radius $ r_h $
\be
U(x) \equiv \frac{v_c(r)}{v_c(r_h)} = \sqrt{x \; \frac{\nu'(\xi)}{\nu'(\xi_h)}} 
\quad , \quad x = \frac{r}{r_h} \; .
\ee
Explicitly solving the Thomas-Fermi eqs.(\ref{nu}) we find
that $ v_c(r)/v_c(r_h) $ is {\bf only} function of $ x = r/r_h $
and takes the same values for all galaxy masses in the range going
from $ 5.13 \;  10^9 \; M_{\odot} $ till $ 5.15 \; 10^{11} \; M_{\odot} $
as shown in fig. \ref{urc}. Namely, $ U(x) $ turns to be an {\bf universal} function.

{\vskip 0.1cm} 

This is a remarkable result since a priori, $ v_c(r)/v_c(r_h) $ could be
a function of $ r $ {\bf and} $ r_h $ and could be different for different
galaxies.

This important result shows the ability of the Thomas-Fermi approach
to correctly describe the galaxy structures.

\bigskip

The phase-space density $ Q(r) $ can be also obtained from the circular
velocity at the radius $ r $ as
\be\label{qv}
 Q_c(r)  = 3 \; \sqrt3 \; \frac{\rho(r)}{v_c^3(r)} \quad , \quad 
Q_{c \, h} = 3 \; \sqrt3 \; \frac{\rho(r_h)}{v_c^3(r_h)} \; .
\ee
Expressing $ \rho(r_h) $ in terms of $ r_h $ eq.(\ref{dilu}),
%and the surface density 
$ \Sigma_0 $ eq.(\ref{densu}) and $ v_c(r_h) $ from eq.(\ref{vtf}) yields
for the phase-space density at the halo radius:
\be\label{qc}
 Q_{c \, h} = 7.96204 \; \frac{I_2(\nu_0)}{\left[-\xi_h \; \nu'(\xi_h)\right]^{\frac32}}
\; \left(\frac{m}{2 \, {\rm keV}}\right)^4 \;  {\rm keV}^4 \; .
\ee
The numerical values of $ Q_{c \, h} $  turn to be larger than $ Q(0) $ eq.(\ref{Qcero}) 
approximately by a factor of two. In the dilute regime $ M_h \gtrsim 10^6 \; M_\odot $ they are related by
\be\label{q0qch}
 Q_{c \, h} = 2.0873 \; Q(0) \; .
\ee

%\newpage

\begin{table*}
\centering
\begin{minipage}{140mm}
\caption{Observed values $ r_h, \; \rho(0) $ 
and $ M_h $ covering from ultracompact galaxies to large spiral galaxies
from the references: \citep{wp,sal07,newa,gil,jdsmg,simon11,wolf10,brodie,willm,martinez,sal12}.
The errors are generally about 20\%-40\% and they are explicitly given when available.}
\begin{tabular}{|c|c|c|c|} \hline  
 & & & \\
 Galaxy  & $ \displaystyle \frac{r_h}{\rm pc} $ 
& $ \rho(0)/\displaystyle \frac{M_\odot}{({\rm pc})^3} $ & $ \displaystyle \frac{M_h}{10^6 \; M_\odot} $
\\ & & &  \\ \hline 
Willman 1 & $ 33^{+7}_{-8} $ & $ 6.8\pm 3 $ & $ 0.39^{+2.5}_{-1.6} $
\\ \hline  
 Segue 1 & $ 52 \pm 10 $ & $ 2.5^{+4.1}_{-1.9} $ & $ 0.58^{+8.2}_{-5.1} $ \\ \hline  
  Leo IV & $ 166 \pm 160 $ & $ .19 \pm 0.2 $ & $ 1.4 \pm 1.5 $ \\ \hline  
Canis Venatici II & $ 145 \pm 70 $ & $ 0.49 \pm 0.25 $ & $ 2.4 \pm 1.1 $
\\ \hline  
Coma-Berenices & $ 109 \pm 40 $ & $ 2.09 \pm 0.86 $  & $ 1.2 \pm 0.4 $
\\ \hline  
 Leo II & $ 57.5^{+166}_{-51} $ & $ 4.07^{+530}_{-3.5} $ & $ 14.1 \pm 0.3 $
\\  \hline  
 Leo T & $ 170 \pm 23 $ & $ 0.79\pm 0.36 $ & $ 12.9 \pm 7 $
\\ \hline  
 Hercules & $ 354 \pm 140 $ & $ 0.1 \pm 0.04 $ & $ 7.1 \pm 2.6 $
\\ \hline  
 Carina & $ 603^{+545}_{-287} $ & $ 0.065^{+0.07}_{-0.03} $ & $ 11.7 \pm 0.3 $
\\ \hline 
 Ursa Major I & $ 458 \pm 150 $ & $ 0.25\pm 0.08 $ & $ 15 \pm 4 $
\\ \hline  
 Draco & $ 646^{+401}_{-266} $ & $ 0.18^{+0.12}_{-0.06} $ & $ 20.4 \pm 0.5 $
\\ \hline  
 Leo I & $ 282^{+255}_{-150} $  & $ 0.41^{+0.94}_{-0.23} $ & $ 38 \pm 0.6 $
\\ \hline  
 Sculptor & $ 355^{+124}_{-80} $ & $ 0.25^{+0.1}_{-0.08}  $ & $ 35.5 \pm 0.3 $
\\ \hline 
 Bo\"otes I & $ 406 \pm 36 $ & $ 0.22^{+0.32}_{-0.12} $ & $ 23.6^{+20}_{-10} $
\\ \hline  
 Canis Venatici I & $ 596 \pm 150 $  & $ 0.08\pm 0.02 $ & $ 27 \pm 4 $
\\ \hline  
Sextans & $ 46.8^{+115}_{-34} $ & $ 5.5^{+130}_{-5.2} $ & $ 32.4 \pm 0.7 $
\\ \hline 
 Ursa Minor & $ 245^{+317}_{-156} $ & $ 0.41^{+1.9}_{-0.3} $ & $ 36.3 \pm 0.5 $
\\ \hline  
 Fornax  & $ 372^{+85}_{-90} $ & $ 0.19 ^{+0.12}_{-0.06} $ & $ 126 \pm 0.06 $
\\  \hline  
 NGC 185  & 450 & $ 4.09 $ & $ 975 \pm 300$
\\ \hline  
 NGC 855  & 1063 & $ 2.64 $ & $ 8340 \pm 1600 $
\\ \hline  
 Small Spiral URC & $ 5100 \pm 1550 $ & $ 0.029 \pm 0.001 $ & $ 6900 \pm 2300 $
\\ \hline  
NGC 4478 & 1890 & $ 3.7 $ & $ 6.55 \times 10^4 \pm 7 \times 10^3 $
\\ \hline  
 Medium Spiral URC & $ 1.9 \times 10^4 \pm 6 \times 10^3 $ & $ 0.0076 \pm 0.002 $ & $ 1.01 \times 10^5 \pm 3 \times 10^4 $
\\ \hline  
 NGC 731 & 6160 & $ 0.47 $ & $ 2.87 \times 10^5 \pm 3 \times 10^4 $
\\ \hline 
 NGC 3853   & 5220 & $ 0.77 $  & $ 2.87 \times 10^5 \pm 4.5 \times 10^4 $ \\ \hline 
NGC 499  & 7700 &  $ 0.91 $ & $ 1.09 \times 10^6 \pm 2 \times 10^5 $ \\   \hline 
Large Spiral URC & $ 5.9 \times 10^4 \pm 1.8 \times 10^3 $ & $ 2.3 \times 10^{-3} \pm 7 \times 10^{-4} $ & 
$ 1. \times 10^6 \pm 3 \times 10^5 $ \\ \hline  
\end{tabular}
\end{minipage}
\label{pgal}
\end{table*}

\section{Circular Velocities contrasted with observations}

The circular velocities values $ v_c(r) $ eq.(\ref{vci}) are known with precision from
galaxy observational data. 

\subsection{The galaxy data}

The kinematics of about several thousands disk galaxies,  
described by the Rotation Curves of Spirals, and the information
obtained from other tracers of the gravitational field of galaxies, 
including the  dispersion velocities of spheroidals and the weak lensing measurements
 (\citealp{sal07} and references therein)
show that the density of the dark matter halos around galaxies of 
different kinds, different luminosity and Hubble types is well represented, 
out to the galaxy virial radius, by an empirical Burkert profile
\bea\label{bur}
&& \rho_{B}(r) =  \rho_{0\; B} \; F_B\left(\frac{r}{r_{h \; B}}\right) \; , 
\cr \cr
&& F_B(x) =  \frac1{(1+x) \; (1 + x^2)} \; , \; x \equiv \frac{r}{r_{h \; B}} \; ,
\eea
where  $ \rho_0 $ stands for the central core density and $ r_{h \; B} $ for
the core radius. The empirical Burkert profile satisfactorily fits the astronomical 
observations and we use the observed data of $ \rho_{0\; B} $ and $ r_{h \; B} $ for 
DM dominated spiral galaxies given in  \citep{sal07}.

\medskip

Kinematical data and properties of other
galaxy gravitational potential tracers are all reproduced,
within their observational uncertainty by a mass model
including a DM halo with a Burkert profile (see \cite{dona,sal07}).
While some other cored DM
distributions [but not the pseudo-isothermal one $ \rho_0/(r^2 + a^2) $]
may succesfully reproduce these data, every cuspy distribution 
fails to do so (see e. g. \cite{gen}).

{\vskip 0.1cm} 

The Burkert and the \cite{jpena} profiles are indistinguishable
with the data available at present.
Any cored density profile with two free parameters (central
density, core radius) that decreases faster than $ 1/r^2 $ can be
mapped into each other, just by a transformation of
the parameters. Only when the determination of the DM
density profile will be available at the
few per cent error level (today it is 10-30 percent) we would be able to 
discriminate between different cored the profiles.

\medskip 

The circular velocities $ V_{URC, \, h}(r) $
for the empirical Burkert density profile follow  
from eq.(\ref{vci}) and eq.(\ref{bur}) \citep{sal07}
\bea\label{vbur}
&& V^2_{URC, \, h}(r)=  2 \, \pi \; G \; \frac{\rho_{0\; B} \; r_{h \; B}^3}{r} \times \cr \cr
&& \! \! \! \! \! \! \! \! \left[ \ln(1+x) -\arctan x + \frac12 \; \ln(1+x^2) \right] \, , \,
 x = \frac{r}{r_{h \; B}}  \; . 
\eea
Notice that normalizing $ V_{URC, \, h}(r) $ to its value at the core radius $ r_{h \; B} $ yields
\bea\label{ubur}
&& U_{URC}^2(x)\equiv \frac{ V^2_{URC, \, h}(r)}{V^2_{URC, \, h}(r_h)} = \cr \cr
&&\! \! \! \! =\frac{3.93201}{x} 
\left[\ln(1+x) -\arctan x + \frac12 \; \ln(1+x^2) \right] \; .
\eea
Namely, the function $ U(x)_{URC} $ only depends on $ x = r/r_{h \; B} $ and complies with 
the concept of Universal Rotation Curve (URC, \citealp{sal07}): $ U(x)_{URC} $
is an {\bf universal} function.

{\vskip 0.1cm} 

Notice that the URC concept is valid not only for the Burkert representation
of the density profile but also for other density profiles that correctly
reproduce the density data.

\medskip

In the Burkert profile case, the halo galaxy mass follows integrating eq.(\ref{bur})
from zero to $ r_{h \; B} $
\be\label{mbur}
M_h = 1.59796 \; \rho_{0\; B} \; r_{h \; B}^3 = 1.59796 \; \Sigma_0 \; r_{h \; B}^2 \; .
\ee
This empirical equation can be recasted in a similar form to eq.(\ref{dilu}) of the 
theoretical Thomas-Fermi approach,
\be \label{burh}
r_{h \; B} = 72.215 \; \sqrt{\frac{120}{\Sigma_0 \;  {\rm pc}^2}\frac{M_h}{\; 10^6}} \; \; {\rm pc}\; .
\ee
$ r_h $ in eqs.(\ref{dilu}) and $ r_{h \; B} $ (\ref{burh}) refer to the point where 
$ \rho(r_h)/\rho(0) = 1/4 $ and $ \rho_{B}(r_{h \; B})/\rho(0) = 1/4 $ both
according to eq.(\ref{onequarter}), for two different density profiles: 
the theoretical Thomas-Fermi profile $ \rho(r) $ in eq.(\ref{dilu})
and the empirical Burkert profile $ \rho(r)_{B} $ for the observational data.

\medskip

\begin{figure}
\begin{turn}{-90}
\psfrag{"Rperfu10.dat"}{$ {\hat M}_h= 7 \; 10^{11} $}
\psfrag{"Rperfu11.dat"}{$ {\hat M}_h= 6.2 \; 10^8 $}
\psfrag{"Rperfu12.dat"}{$ {\hat M}_h= 1.3 \; 10^8 $}
\psfrag{"Rperfu13.dat"}{$ {\hat M}_h= 2.5 \; 10^7 $}
\psfrag{"Rperfu14.dat"}{$ {\hat M}_h= 5.1 \; 10^6 $}
\psfrag{"Rperfu15.dat"}{$ {\hat M}_h= 1.1 \; 10^6 $}
\psfrag{"Rperfu16.dat"}{$ {\hat M}_h= 2.2 \; 10^5 $}
\psfrag{"Rperfu30.dat"}{$ {\hat M}_h= 1.6 \; 10^5 $}
\psfrag{"RperfuB.dat"}{Universal Burkert profile}
\psfrag{equis}{$ x = r/r_h $}
\psfrag{rhorrho0}{$ \rho(r)/\rho(0) $}
\includegraphics[height=9.cm,width=9.cm]{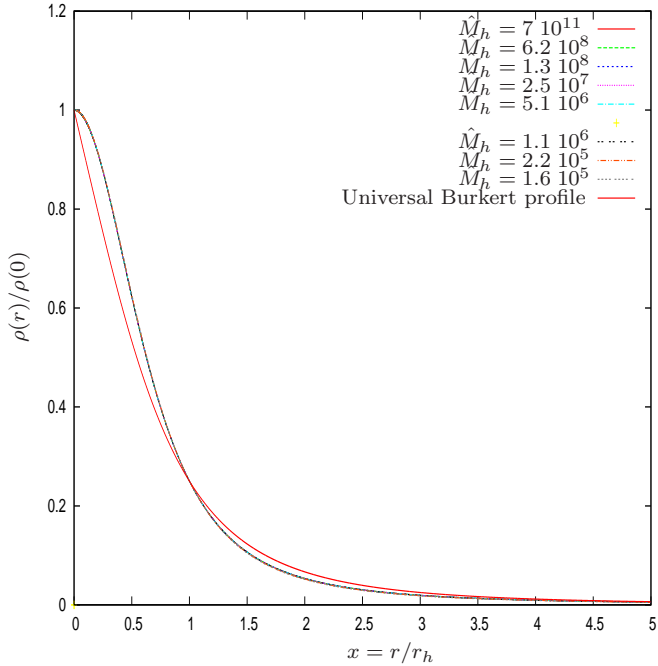}
\end{turn}
\caption{Normalized density profiles $ \rho(r)/\rho(0) $ as functions of $ r/r_h $.
We display the theoretical profiles for galaxy masses in the dilute regime
$  1.4 \; 10^5 < {\hat M}_h < 7.5 \; 10^{11} , \; -1.5 > \nu_0 > -20.78 $. {\bf All} fall into
the {\bf same and universal} density profile. We plot the empirical Burkert profile as function of $ r/r_h $.}
\label{perfus}
\end{figure}

The halo radius $ r_{h \; B} $ and $ r_h $ for given galaxy mass $ M_h $ and 
surface density $ \Sigma_0 $ are related by the universal relation
$$
 r_h = 0.95401 \;  r_{h \; B} \; ,
$$
that follows from eqs.(\ref{dilu}) and (\ref{mbur}). Namely, due to the slight shape
difference between the theoretical Thomas-Fermi and empirical Burkert profiles (see fig. \ref{perfus}), the 
Thomas-Fermi halo radius turns to be about 5\% smaller than the Burkert halo radius.

\medskip

It follows from eq.(\ref{nu}) that
the theoretical Thomas-Fermi profile posses an expansion in even powers of $ r^2 $
(this is also the case for the density profiles obtained in the linear approximation
from the cosmological density fluctuations in \cite{newa}).
On the contrary, the empirical Burkert profile eq.(\ref{bur}) is not an even function of $ r $
and exhibits a linear behaviour in $ r $ near the origin. This is the source
of the small deviation  near $ r = 0 $ between the theoretical Thomas-Fermi profile and 
the empirical Burkert profile exhibited in fig. \ref{perfus}.

\medskip

The circular velocity at the halo radius 
for the empirical Burkert profile follows setting $ x = 1 $ in eq.(\ref{vbur}) with the result
$$
V_{URC, \, h}(r_h)= 7.717 \; \left(\frac{M_h}{10^6 \;  M_\odot}\right)^{\! \! \frac14} \;
\left(\frac{\Sigma_0 \;  {\rm pc}^2}{120 \; M_\odot}\right)^{\! \! \frac14}  \; 
\; \frac{\rm km}{\rm s} \; .
$$
This value is to be compared with the theoretical Thomas-Fermi result
eq.(\ref{vcrh}). We see that they differ from each other by only 2.4 \%,
confirming again the success of the  Thomas-Fermi approach to
describe the galaxy structures.

\begin{figure}
\begin{turn}{-90}
\psfrag{"Fvcirc1.dat"}{$ M_h = 5.1 \; 10^9 \; M_{\odot} $: Theory}
\psfrag{"vcsalu1.dat"}{Observational}
\psfrag{"Fvcirc2.dat"}{$ M_h = 8.4 \; 10^9\; M_{\odot} $: Theory}
\psfrag{"vcsalu2.dat"}{Observational}
\psfrag{"Fvcirc3.dat"}{$ M_h = 1.4 \; 10^{10}\; M_{\odot} $: Theory}
\psfrag{"vcsalu3.dat"}{Observational}
\psfrag{"Fvcirc4.dat"}{$ M_h = 2.3 \; 10^{10}\; M_{\odot} $: Theory}
\psfrag{"vcsalu4.dat"}{Observational}
\psfrag{"Fvcirc5.dat"}{$ M_h = 3.8 \; 10^{10} \; M_{\odot} $: Theory}
\psfrag{"vcsalu5.dat"}{Observational}
\psfrag{equis}{$ r $ in kpc}
\psfrag{vcr}{$ v_c(r) $ in km/s}
\includegraphics[height=9.cm,width=9.cm]{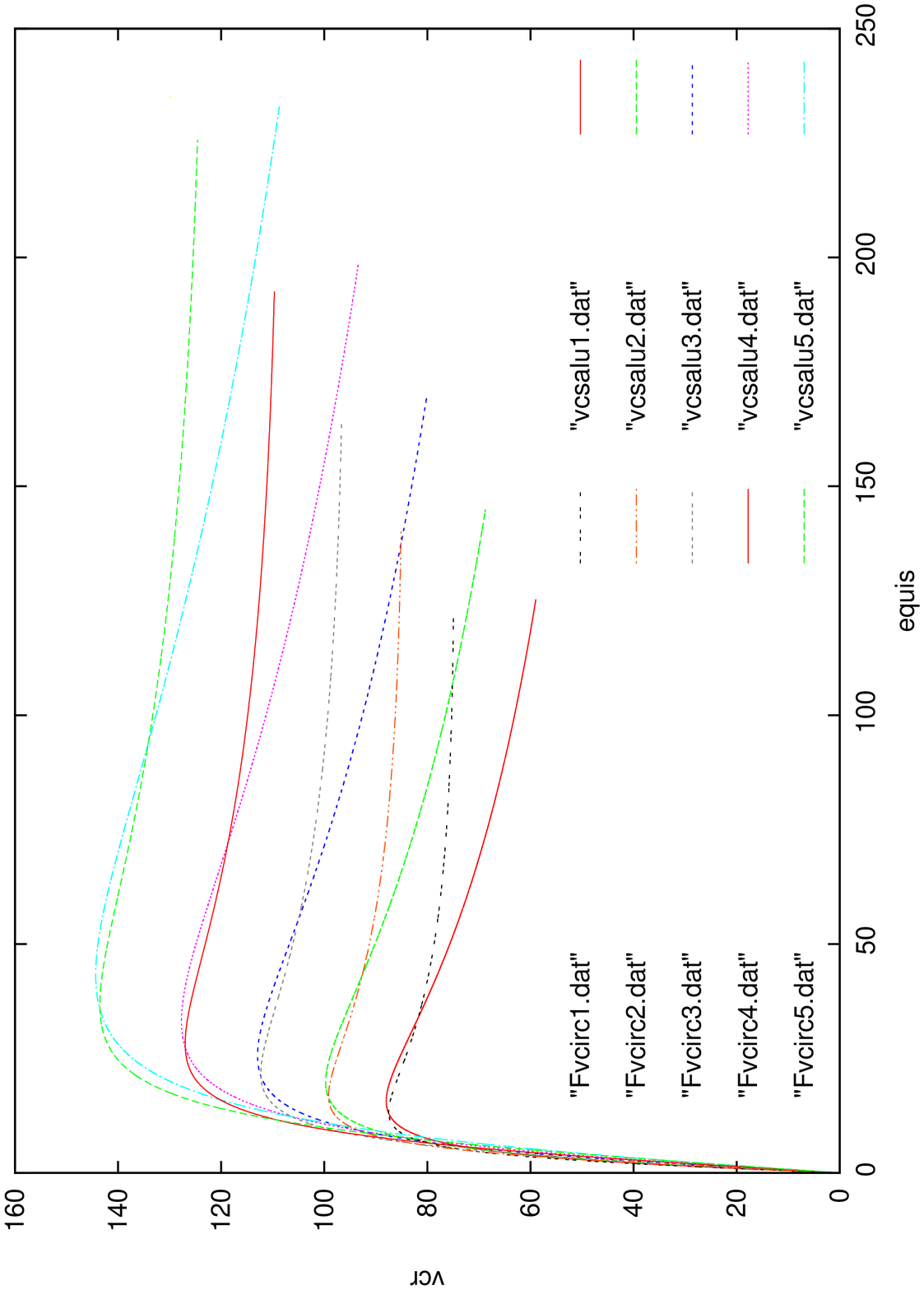}
\psfrag{"Fvcirc6.dat"}{$ M_h = 6.4 \; 10^{10} \; M_{\odot} $: Theory}
\psfrag{"vcsalu6.dat"}{Observational}
\psfrag{"Fvcirc7.dat"}{$ M_h = 1.1 \; 10^{11}\; M_{\odot} $: Theory}
\psfrag{"vcsalu7.dat"}{Observational}
\psfrag{"Fvcirc8.dat"}{$ M_h = 1.8 \; 10^{11}\; M_{\odot} $: Theory}
\psfrag{"vcsalu8.dat"}{Observational}
\psfrag{"Fvcirc9.dat"}{$ M_h = 3.0 \; 10^{11}\; M_{\odot} $: Theory}
\psfrag{"vcsalu9.dat"}{Observational}
\psfrag{"Fvcirc10.dat"}{$ M_h = 5.2 \; 10^{11} \; M_{\odot} $: Theory}
\psfrag{"vcsalu10.dat"}{Observational}
\psfrag{equis}{$ r $ in kpc}
\psfrag{vcr}{$ v_c(r) $ in km/s}
\includegraphics[height=9.cm,width=9.cm]{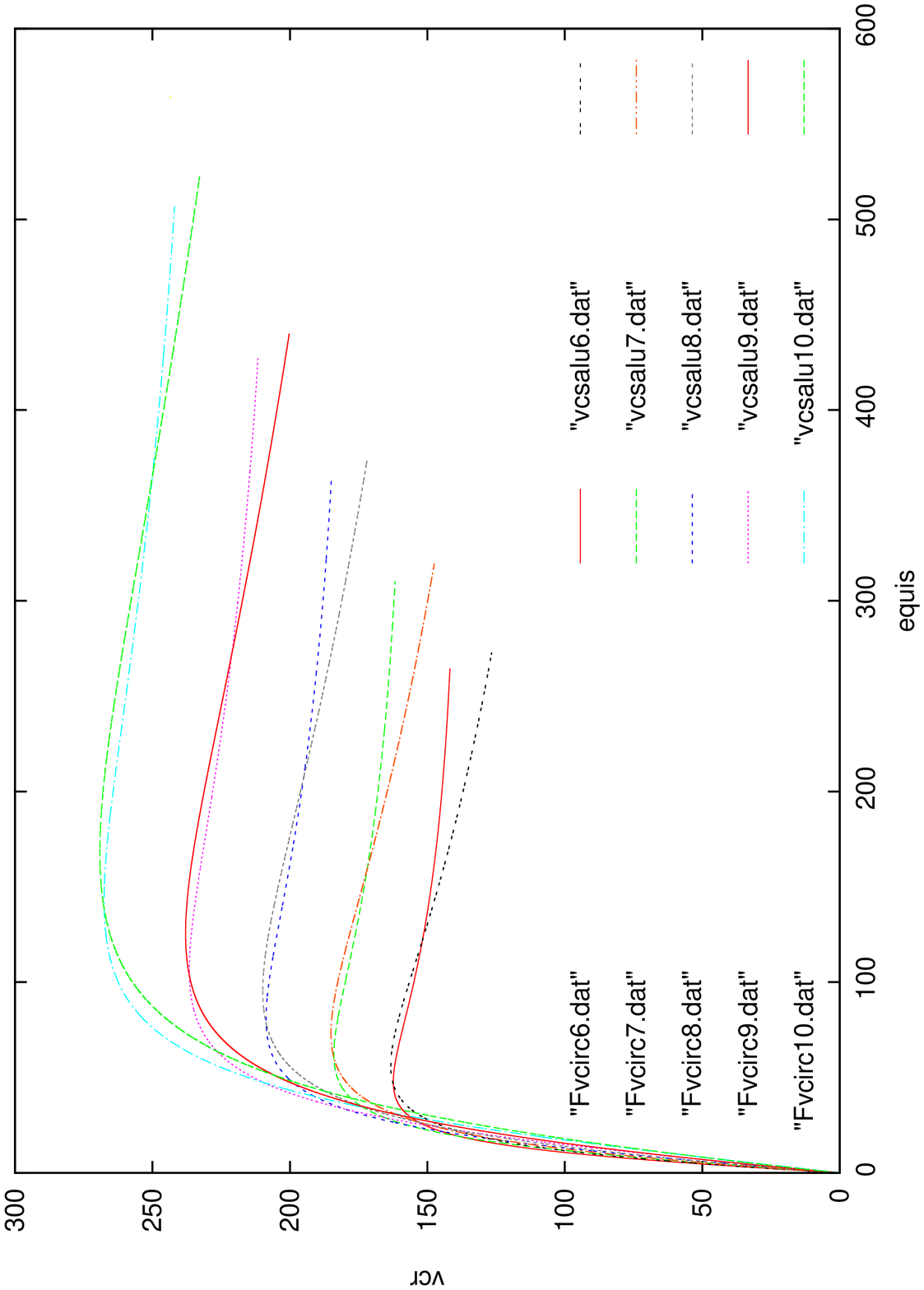}
\end{turn}
\caption{The velocity rotation curves $ v_c(r) $ in km/s versus $ r $ in kpc
for ten different independent galaxy masses $ M_h $ going from $ 5.13 \;  10^9 \; M_{\odot} $
till $ 5.15 \; 10^{11} \; M_{\odot} $. For each galaxy mass  $ M_h $,
we show the two curves: the theoretical Thomas-Fermi curve and the observational
curve described by the empirical Burkert profile. The Thomas-Fermi curves reproduce 
remarkably well the observational curves for $ r \lesssim r_h $.  
We plot $ v_c(r) $ for $ 0 < r < r_{vir} , \; r_{vir} $
being the virial radius of the galaxy.}
\label{vcirc}
\end{figure}

\subsection{Comparison to observations}

\begin{figure}
\begin{turn}{-90}
\psfrag{"perfd1.dat"}{$ M_h = 5.1 \; 10^9 \; M_{\odot} $: Theory}
\psfrag{"4perfB1.dat"}{Observational}
\psfrag{"perfd7.dat"}{$ M_h = 1.1 \; 10^{11}\; M_{\odot} $: Theory}
\psfrag{"4perfB7.dat"}{Observational}
\psfrag{equis}{$ r $ in kpc}
\psfrag{logrho}{$ \log_{10}[\rho(r)/(M_\odot/{\rm pc}^3)] $}
\includegraphics[height=9.cm,width=9.cm]{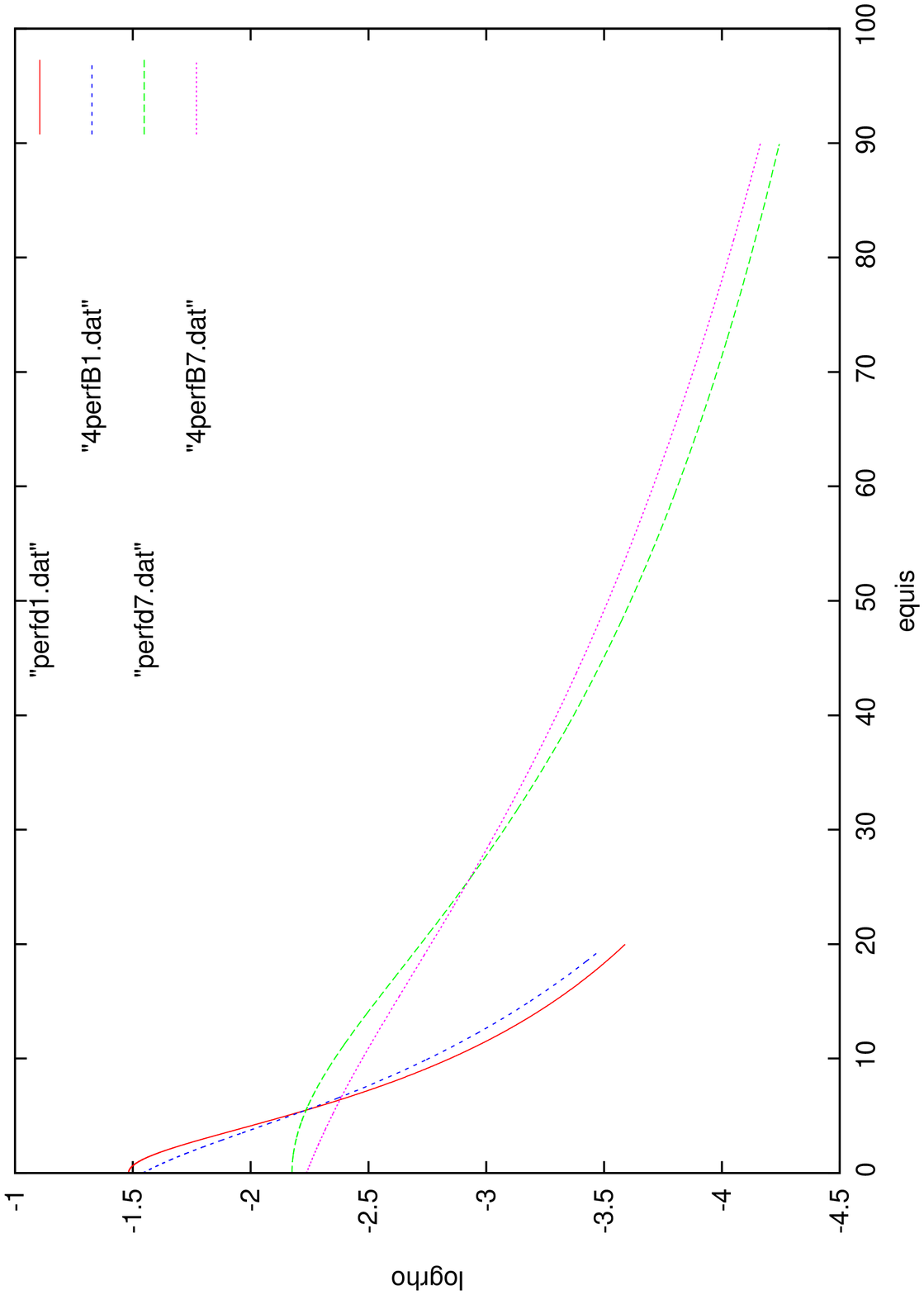}
\psfrag{"perfd2.dat"}{$ M_h = 8.4 \; 10^9\; M_{\odot} $: Theory}
\psfrag{"4perfB2.dat"}{Observational}
\psfrag{"perfd5.dat"}{$ M_h = 3.8 \; 10^{10} \; M_{\odot} $: Theory}
\psfrag{"4perfB5.dat"}{Observational}
\psfrag{"perfd8.dat"}{$ M_h = 1.8 \; 10^{11}\; M_{\odot} $: Theory}
\psfrag{"4perfB8.dat"}{Observational}
\psfrag{equis}{$ r $ in kpc}
\psfrag{logrho}{$ \log_{10}[\rho(r)/(M_\odot/{\rm pc}^3)] $}
\includegraphics[height=9.cm,width=9.cm]{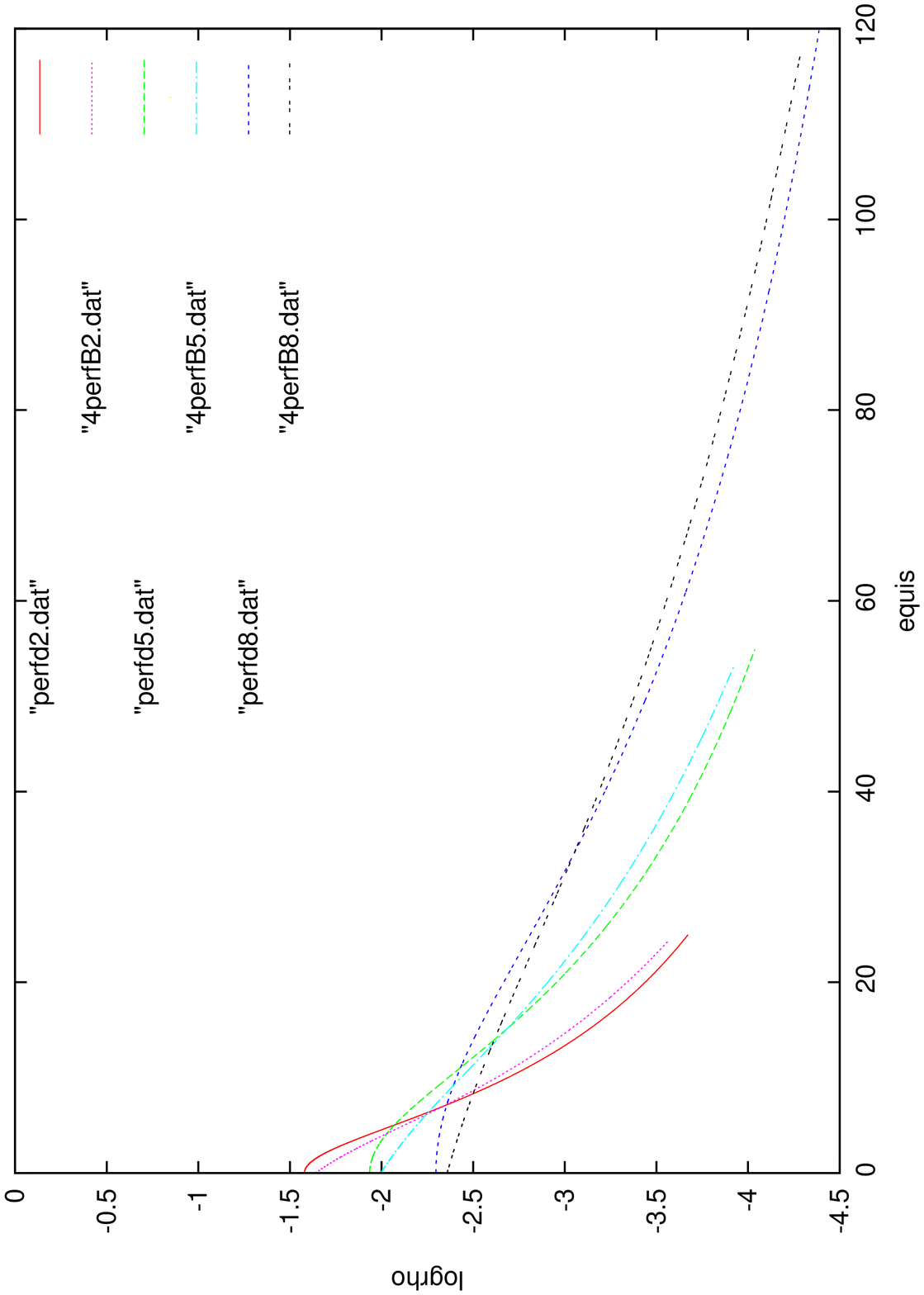}
\end{turn}
\caption{The theoretical Thomas-Fermi density profiles 
and the observational profiles described by the Burkert expression
for the first five galaxy masses. We plot the ordinary logarithm of the density
in $ M_\odot/{\rm pc}^3 $ vs. $ r $ in kpc in the interval $ 0 < r < 4 \; r_h $.
For each galaxy mass  $ M_h $, we show the two curves: the theoretical 
Thomas-Fermi curve and the observational Burkert curve.
The agreement of the Thomas-Fermi curves to the observational curves
is remarkable.}
\label{perfi1}
\end{figure}

\begin{figure}
\begin{turn}{-90}
\psfrag{"perfd3.dat"}{$ M_h = 1.4 \; 10^{10}\; M_{\odot} $: Theory}
\psfrag{"4perfB3.dat"}{Observational}
\psfrag{"perfd6.dat"}{$ M_h = 6.4 \; 10^{10} \; M_{\odot} $: Theory}
\psfrag{"4perfB6.dat"}{Observational}
\psfrag{"perfd9.dat"}{$ M_h = 3.0 \; 10^{11}\; M_{\odot} $: Theory}
\psfrag{"4perfB9.dat"}{Observational}
\psfrag{equis}{$ r $ in kpc}
\psfrag{logrho}{$ \log_{10}[\rho(r)/(M_\odot/{\rm pc}^3)] $}
\includegraphics[height=9.cm,width=9.cm]{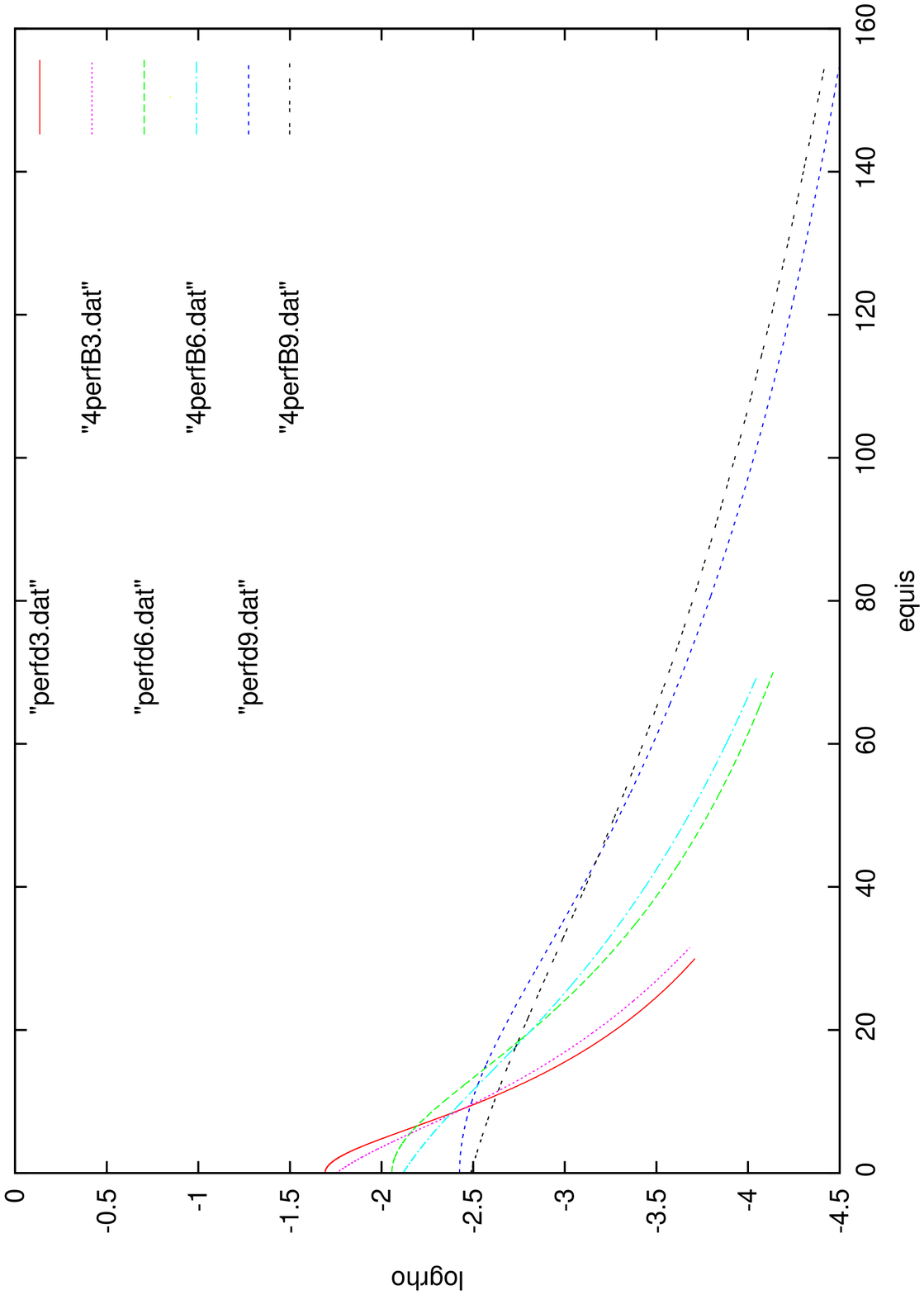}
\psfrag{"perfd4.dat"}{$ M_h = 2.3 \; 10^{10}\; M_{\odot} $: Theory}
\psfrag{"4perfB4.dat"}{Observational}
\psfrag{"perfd10.dat"}{$ M_h = 5.2 \; 10^{11} \; M_{\odot} $: Theory}
\psfrag{"4perfB10.dat"}{Observational}
\psfrag{equis}{$ r $ in kpc}
\psfrag{logrho}{$ \log_{10}[\rho(r)/(M_\odot/{\rm pc}^3)] $}
\includegraphics[height=9.cm,width=9.cm]{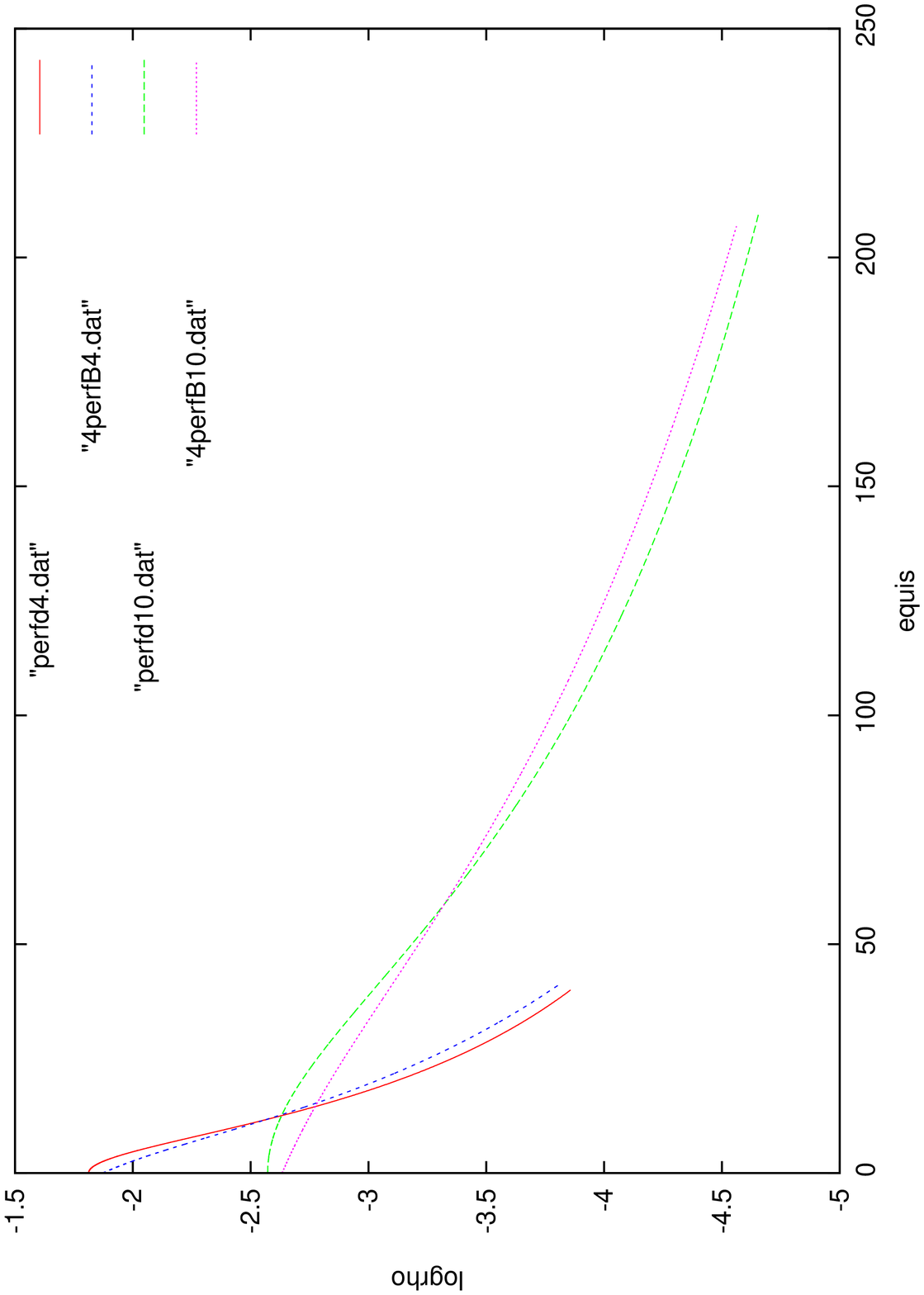}
\end{turn}
\caption{The theoretical Thomas-Fermi density profiles 
and the observational profiles described by the Burkert expression
for further five galaxy masses. We plot the ordinary logarithm of the density
in $ M_\odot/{\rm pc}^3 $ vs. $ r $ in kpc in the interval $ 0 < r < 4 \; r_h $.
For each galaxy mass  $ M_h $, we show the two curves: the theoretical 
Thomas-Fermi curve and the observational Burkert curve.
The agreement of the Thomas-Fermi curves to the observational curves
is remarkable.}
\label{perfi2}
\end{figure}

Our results are {\bf independent} of the details of the WDM particle physics model.
They follow from the gravitational self-interaction
of WDM particles and their fermionic nature.
The same remarks apply to all the Thomas-Fermi results 
including the lower bound $ m > 1.91 $ keV \citep{astro}.

\medskip

We depict in fig. \ref{urc} the normalized circular velocities
$ U(x) = v_c(r)/ v_c(r_h) $ vs. $ x = r/r_h $ obtained on one hand
from the observational data $ U(x)_{URC} $ described with the empirical
Burkert
profile eq.(\ref{ubur}), and on the other hand, $ U(x) $ obtained from the
theoretical Thomas-Fermi formula eq.(\ref{vtf}).
In general, the normalized circular velocities $ v_c(r)/ v_c(r_h) $
can be functions of $ r $ and $ r_h $, which just reflects the fact
that in a spherically symmetric approach only one parameter $ r_h $ shows up in the circular velocities. 
[The parameter $ r_h $ depends on the characteristic
energy $ E_0 $ in the Fermi-Dirac distribution function at fixed surface
density $ \Sigma_0 $. Generalizing the Thomas-Fermi approach
to non-spherically symmetric and non-isotropic situations
by including other particle parameters like the angular momentum
in the distribution functions would lead to density profiles 
depending on other parameters besides $ r_h $.]

Here we remarkably find that the normalized
circular velocities
$ v_c(r)/ v_c(r_h) $ turn out to be functions of {\bf only} one variable:
the ratio $ x = r/r_h $. [On the contrary, $ v_c(r) $ is a
function of $ r $ {\bf and} $ r_h $ separately].
We see that the theoretical curves from the Thomas-Fermi approach for ten
different galaxy masses
{\bf all fall one into each other}. Therefore, we find the result that the
Thomas-Fermi approach
provides {\bf universal} rotation curves. Moreover,  the theoretical 
Thomas-Fermi curves $ U(x) $
and  the observational
universal curve $ U(x)_{URC} $ described by the empirical Burkert 
profile {\bf coincide} for $ r < r_h $.

\medskip

We depict in fig. \ref{perfus} the 
normalized density profiles $ F(x) = \rho(r)/\rho(0) $ as functions of $ x = r/r_h $
obtained from the theoretical Thomas-Fermi profiles for galaxy masses in the dilute regime
$  1.4 \; 10^5 < {\hat M}_h < 7.5 \; 10^{11} , \; -1.5 > \nu_0 > -20.78 $. {\bf All} fall into
the {\bf same and universal} density profile. The empirical Burkert profile $ F_B(x) $ 
in fig. \ref{perfus} turns to be very close to the theoretical Thomas-Fermi profile $ F(x) $ 
except near the origin as discussed above.

\medskip

We display in figs. \ref{vcirc} $ v_c(r) $ in km/s vs. $ r $ in kpc obtained
on one hand from the observational data described with the empirical Burkert profile
eq.(\ref{vbur}) 
and on the other hand from the theoretical Thomas-Fermi formula eq.(\ref{vtf}).
We plot in figs. \ref{vcirc} $ v_c(r) $ for $ 0 < r < r_{vir} , \; r_{vir} $
being the virial radius of the galaxy.

{\vskip 0.1cm} 

The corresponding halo galaxy masses $ M_h $ are indicated in figs. \ref{vcirc} 
and run from $ 5.13 \; 10^9 M_\odot $ till $ 5.15 \; 10^{11} M_\odot $.

{\vskip 0.1cm} 

The theoretical rotation curves reproduce the observational
curves modelized with the empirical Burkert profile
for $ r \lesssim r_h $ justifying the use of the 
Fermi--Dirac distribution function eq.(\ref{FD})
in the Thomas-Fermi eqs.(\ref{dfI})-(\ref{nu}).

{\vskip 0.2cm} 

We display in figs. \ref{perfi1} and \ref{perfi2}
the theoretical density profiles computed from the Thomas-Fermi
equations and the observational profiles described by the empirical Burkert expression.
We plot the ordinary logarithm of the density
in $ M_\odot/{\rm pc}^3 $ vs. $ r $ in kpc in the interval $ 0 < r < 4 \; r_h $.
We see a very good agreement of the theoretical density profiles
with the observations modelized with the empirical Burkert profile
in all the range $ 0 < r < 4 \; r_h $.

\medskip

\begin{figure}
\begin{turn}{-90}
\psfrag{"CLMLrhmu120.dat"}{observed values}
\psfrag{"Mr0.dat"}{theory curve}
\psfrag{logMh}{$ \log_{10}{\hat M}_h $}
\psfrag{logrh}{$ \log_{10}{\hat r}_h $}
\includegraphics[height=9.cm,width=9.cm]{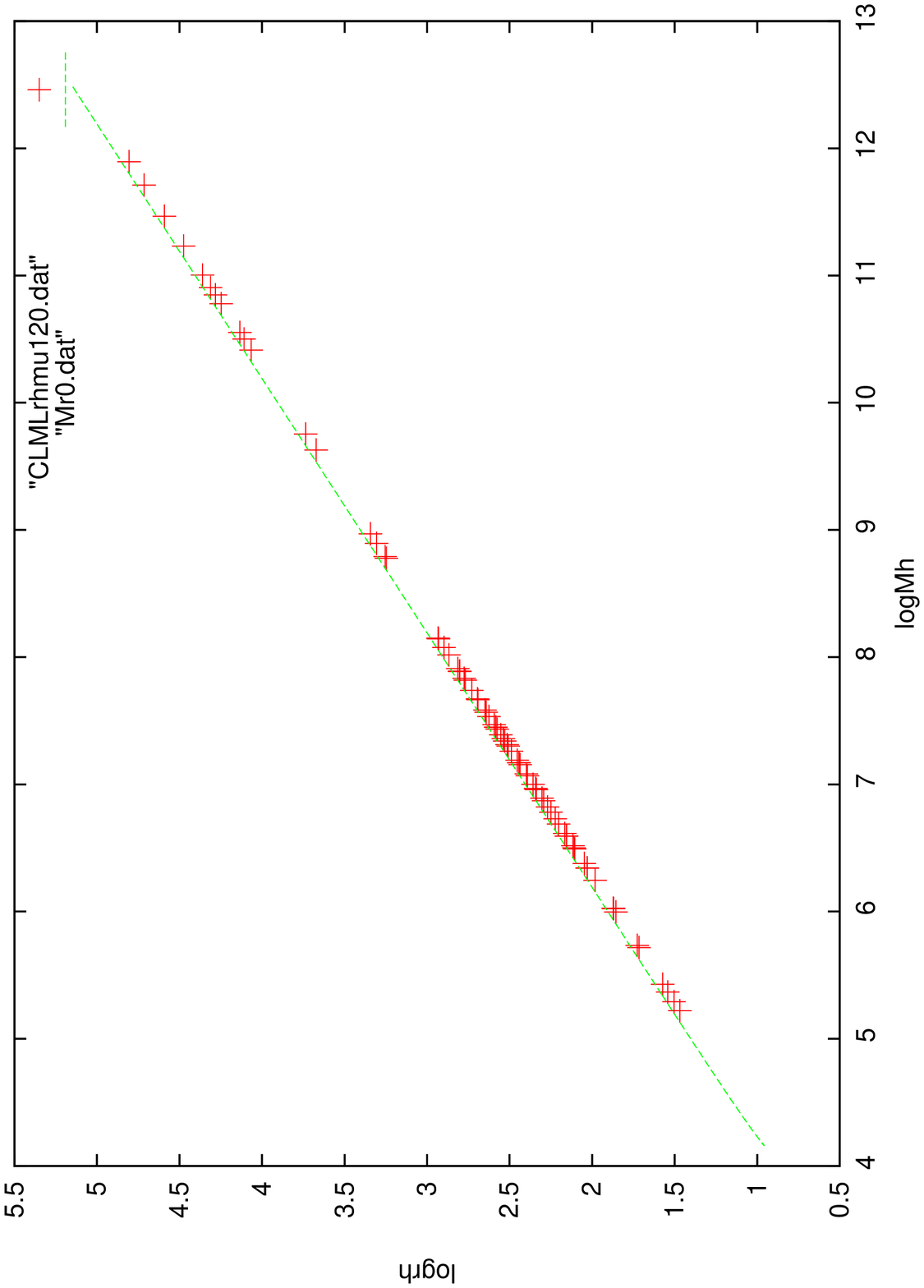}
\end{turn}
\caption{The ordinary logarithm of $ {\hat r}_h = \frac{r_h}{\rm pc} \; \left(\frac{\Sigma_0 \;  
{\rm pc}^2}{120 \; M_\odot}\right)^{\! \! \frac15} $ vs. the ordinary logarithm of 
$ {\hat M}_h = \frac{M_h}{M_\odot} \; \left(\frac{120 \; M_\odot}{\Sigma_0 \;  
{\rm pc}^2}\right)^{\! \! \frac35} $. We see that $ r_h $ follows with precision 
the square-root of $ M_h $ as in the dilute regime eq.(\ref{dilu}) of the Thomas-Fermi equations. 
The data for $ M_h $ and $ r_h $ are taken from Table \ref{pgal},
from \citep{mcc} and from \citep{sal07,sal09} and they are extremely well
reproduced by the theoretical Thomas-Fermi curve.}
\label{mhrh}
\end{figure}

In fig. \ref{mhrh} we plot the ordinary logarithm of
\be\label{rsom}
{\hat r}_h \equiv \frac{r_h}{\rm pc} \; \left(\frac{m}{2 \, {\rm keV}}\right)^{\! \! \frac85} \; 
\left(\frac{\Sigma_0 \;  {\rm pc}^2}{120 \; M_\odot}\right)^{\! \! \frac15} 
\ee
versus the ordinary logarithm of
\be\label{Msom}
{\hat M}_h \equiv \frac{M_h}{M_\odot}  \; \left(\frac{m}{2 \, {\rm keV}}\right)^{\! \! \frac{16}5} \; 
\left(\frac{120 \; M_\odot}{\Sigma_0 \; {\rm pc}^2}\right)^{\! \! \frac35}  \; .
\ee
From eqs.(\ref{dilu}), (\ref{rsom}) and (\ref{Msom}), we find the {\bf scaling} relation
\be\label{sombr}
\log_{10} {\hat r}_h = \frac12 \; \log_{10} {\hat M}_h - 1.16182 \; ,
\ee
which is accurate for $ M_h \gtrsim 10^6 \;  M_\odot $.
We see from fig. \ref{mhrh} that $ {\hat r}_h $ follows with precision the square-root 
law eq. (\ref{sombr}) obtained from the dilute regime eq.(\ref{dilu}) of the Thomas-Fermi equations. 

{\vskip 0.1cm}

For $ m $ in the keV scale, namely $ m \sim 2 $ keV,  we obtain galaxies as 
nondegenerate solutions of the Thomas-Fermi equations with core radius from 
10 pc to 1 Mpc as shown in fig. \ref{mhrh}. Therefore, fermionic WDM 
remarkably well explains the observations of the Fornax and Sculptor dSphs \citep{wp,jpena}

{\vskip 0.1cm} 

Actually, the dilute regime formulas (\ref{dilu})-(\ref{qh}) apply even near the fermion degenerate limit
as shown by fig. \ref{Rmhrh}. In fig. \ref{Rmhrh} we depicted $ \log_{10} {\hat r}_h $ vs. $ \log_{10} {\hat M}_h $
for the smaller galaxies $ {\hat M}_{h, \, min} \leq {\hat M}_h < 10^5 $ where $ {\hat M}_{h, \, min} = 30999 $.
We see that the dilute regime eqs.(\ref{dilu})-(\ref{qh}) reproduce the Thomas-Fermi results for
practically {\bf all} galaxy masses even near the degenerate limit.

{\vskip 0.1cm} 

We plot in fig. \ref{q} the ordinary logarithm of the theoretical phase-space density
$ Q_{c \, h}/{\rm keV}^4 $ vs. the ordinary logarithm of $ {\hat M}_h $
and the observational values of $ \log_{10} Q_{c \, h \, B}/{\rm keV}^4 $. 
We see that the theoretical phase-space density $ Q_{c \, h} $
reproduces very well the observational data parametrized with the empirical Burkert profile.

\medskip

The errors of the data can be estimated to be about 10-20 \%.
The bibliographical sources are \cite{wp,sal07,sal09,newa,gil,jdsmg,simon11,wolf10,brodie,willm,martinez}.

\medskip 

Baryons represent less than 5\% of the galaxy mass \citep{memo,oh,pers}. For
dwarf galaxies baryons count for less than 0.01\% of the galaxy mass
\citep{brodie,martin,mwal,willm,woo}.

{\vskip 0.1cm} 

The self-gravity of the baryonic material is negligible while
baryons are immersed in a DM halo potential well. 
Baryons trace the DM potential well playing the role of test
particles to measure the local DM density. 
In fig. \ref{q} we contrast theory against DM observational data 
gathered from baryons and described with the empirical Burkert profile.

\section{Conclusions}

The more appropriate way to decipher the nature of the dark matter is to study
the properties of the physical objects formed by it: galaxies are formed
overwhelmly by dark matter since 95\% to 99.99 \% of their mass is dark. 
This is the task we pursue in the present paper.

{\vskip 0.1cm} 

Fermionic WDM by itself produce galaxies and structures in 
agreement with observations modelized with the empirical
Burkert profile showing that baryonic corrections to WDM
are not very important.
Therefore, the effect of including baryons is expected to be a
correction to the pure WDM results, consistent with the fact that
dark matter is in average six times more abundant than baryons.

\medskip

The theoretical curves from the Thomas-Fermi approach to galaxy structure for self-gravitating 
fermionic WDM \citep{newas,astro} practically coincide with the
observed galaxy rotation curves and density profiles described with the empirical
Burkert profile for $ r < 2 \; r_h $. In addition, our approach 
provides scaling relations for the main galaxy magnitudes eqs.(\ref{dilu})-(\ref{qh})
as the halo radius $ r_h $, 
mass $ M_h $ and phase space density well in agreement with the observational data.
The set of data is displayed in Table \ref{pgal}.

Therefore, the Fermi-Dirac distribution applies in the region $ r \lesssim 2 \; r_h $
for the whole range of galaxy masses.

{\vskip 0.1cm} 

\begin{figure}
\begin{turn}{-90}
\psfrag{"Cmqv.dat"}{Thomas-Fermi $ \log_{10}  Q_{c \, h}/{\rm keV}^4 $}
\psfrag{"Cqn.dat"}{$ Q_{c \, h \; B} $ Galaxy Data 1}
\psfrag{"qawmcc.dat"}{$ Q_{c \, h \; B} $ Galaxy Data 2}
\psfrag{"Cqpaolo.dat"}{$ Q_{c \, h \; B} $ Galaxy Data 3}
\psfrag{logMh}{$ \log_{10}{\hat M}_h $}
\psfrag{logQh}{$ \log_{10}Q_{c \, h}/{\rm keV}^4 $}
\includegraphics[height=9.cm,width=9.cm]{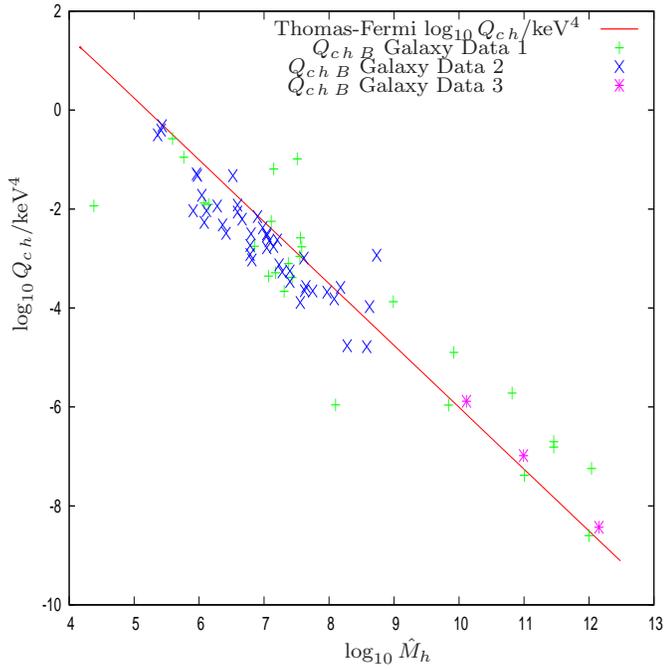}
\end{turn}
\caption{The theoretical $ \log_{10}  Q_{c \, h}/{\rm keV}^4 $
vs. the halo mass $ \log_{10} {\hat M}_h $. The theoretical curve $ Q_{c \, h} $ is obtained
from the Thomas-Fermi expression (\ref{qc}).
The data for $ Q_{c \, h \; B} $ have been obtained from circular velocities.
Galaxy Data 1 refers to data from Table \ref{pgal},
Galaxy Data 2 refers to data from \citep{mcc} and Galaxy Data 3 refers to spiral galaxy data from 
\citep{sal07,sal09} .}
\label{q}
\end{figure}

Notice that (i) the scaling relations eqs.(\ref{dilu})-(\ref{qh}) 
are a consequence solely of the self-gravitating
interaction of the fermionic WDM and (ii) the proportionality factors in these
scaling relations are confirmed by the galaxy data (see figs. \ref{mhrh} and \ref{q}).

{\vskip 0.1cm} 

The galaxy relations derived in eqs.(\ref{dilu})-(\ref{qh}) are accurate for $ M_h \gtrsim 10^6 \;  M_\odot $.
We see that they exhibit a scaling behaviour for $ r_h $ vs. $ M_h $,  
$ Q(0) $ vs. $ M_h $ and $ M_h $ vs. the fugacity at the center $ z_0 = e^{\mu(0)/E_0} $. These scaling behaviours
of the dilute classical regime are {\bf very accurate} even near the degenerate limit as 
shown by fig. \ref{Rmhrh}.
Interestingly enough, the small deviation of these scaling laws 
near the degenerate limit is a manifestation of the quantum effects present in compact dwarf galaxies.

{\vskip 0.1cm} 

The dimensionless halo radius $ {\hat r}_h $ vs. the dimensionless halo mass $ {\hat M}_h $ plotted
in fig. \ref{mhrh} follows with precision the square-root scaling
eq.(\ref{sombr}) obtained from the Thomas-Fermi equations (\ref{pois})
in the dilute regime. Moreover, the observational data 
modelized with the empirical Burkert profile
for $ r_h $ vs. $ M_h $ for a large variety of galaxies
(Table \ref{pgal}, galaxy data in \citealp{mcc}, \citealp{sal07} and \citealp{sal09})
are {\bf all} satisfactorily reproduced by the theoretical Thomas-Fermi curve.

{\vskip 0.1cm} 

Also, as shown by fig. \ref{q} the observational values of the phase-space density 
modelized with the empirical Burkert profile are well 
reproduced by  the theoretical Thomas-Fermi results.

{\vskip 0.1cm} 

The theoretical circular velocities $ v_c(r) $ and the theoretical density profiles $ \rho(r) $
computed from the Thomas-Fermi equations (\ref{pois}) reproduce very well the observational
curves modelized with the empirical Burkert profile
for $ r \lesssim r_h $ as shown in figs. \ref{perfus}, \ref{vcirc},  \ref{perfi1} and \ref{perfi2}.
These results fully justify the use of the 
Fermi--Dirac distribution function in the Thomas-Fermi equations (\ref{pois}).

\medskip

Remarkably enough, solving the Thomas-Fermi eqs.(\ref{pois}) we find that the theoretical
circular velocities $ U(x) = v_c(r)/v_c(r_h) $ as well as the normalized density profiles
$ F(x) =  \rho(r)/\rho(0) $ are {\bf only} functions of $ x = r/r_h $
and take respectively the same value for {\bf all} galaxy masses in the range going
from $ 5.13 \times 10^9 \; M_{\odot} $ till $ 5.15 \times 10^{11} \; M_{\odot} $
as shown in figs. \ref{urc} and \ref{perfus}. Namely, the Thomas-Fermi approach
provides {\bf universal} functions $ U(x) $ and $ F(x) $ for the normalized
circular velocities and normalized density profiles, respectively. 
Moreover,  figs. \ref{urc}  and \ref{perfus} show that the observational universal curves
and the theoretical Thomas-Fermi curves {\bf coincide } for $ r \lesssim 2 \; r_h $.

\medskip

These important results show the ability of the Thomas-Fermi approach
to correctly describe the galaxy structures.

\section*{acknowledgments}

P. S. thanks the Observatoire de Paris LERMA and the CIAS for their kind invitation and hospitality.

\label{lastpage}

\begin{thebibliography}{99}
\bibitem[\protect\citeauthoryear{Anderhalden et al.}{2013}]{ander}
Anderhalden D. et al. arXiv:1212.2967, JCAP, 03, 014 (2013).
\bibitem[\protect\citeauthoryear{Avila-Reese et al.}{2001}]{avi}
V. Avila-Reese et al., Ap J, 559, 516 (2001).
\bibitem[\protect\citeauthoryear{Brodie  et al.}{2011}]{brodie}
J. P. Brodie  et al., AJ, 142, 199 (2011).
 \bibitem[\protect\citeauthoryear{Col\'{\i}n et al.}{2000}]{colin}
P. Col\'{\i}n, O. Valenzuela, V.  Avila-Reese, Ap J, 542, 622 (2000).
\bibitem[\protect\citeauthoryear{Col\'{\i}n et al.}{2008}]{colin8}
P. Col\'{\i}n, O. Valenzuela, V.  Avila-Reese, Ap J, 673, 203 (2008).
\bibitem[\protect\citeauthoryear{Cosmic Frontier}{2013}]{cosmf}
Cosmic Frontier, H. J. de Vega, N. G. Sanchez arXiv:1304.0759.
\bibitem[\protect\citeauthoryear{DdVS}{2013a}]{newas}
C. Destri, H. J. de Vega, N. G. Sanchez, New Astronomy {\bf 22}, 39 (2013).
\bibitem[\protect\citeauthoryear{DdVS}{2013b}]{astro}
C. Destri, H. J. de Vega, N. G. Sanchez, Astrop. Phys., {\bf 46}, 14 (2013).
\bibitem[\protect\citeauthoryear{DdVS}{2013c}]{fluct}
C. Destri, H. J. de Vega, N. G. Sanchez, Phys. Rev. D88, 083512 (2013).
\bibitem[\protect\citeauthoryear{de Vega \& Sanchez}{2010}]{mnras}
H. J. de Vega, N. G. S\'anchez, Mon. Not. R. Astron. Soc. {\bf 404}, 885 (2010).
\bibitem[\protect\citeauthoryear{dVSS}{2010}]{newa}
H. J. de Vega,  P. Salucci, N. G. Sanchez, New Astronomy {\bf 17}, 653 (2012).
\bibitem[\protect\citeauthoryear{Donato et al.}{2009}]{dona} 
F. Donato et al., MNRAS {\bf 397}, 1169 (2009).
\bibitem[\protect\citeauthoryear{Gao \& Theuns}{2007}]{theuns}
L. Gao and T. Theuns,  Science, 317, 1527 (2007).
\bibitem[\protect\citeauthoryear{Gentile et al.}{2004}]{gen}
G. Gentile et al.,  MNRAS 351, 903 (2004).
\bibitem[\protect\citeauthoryear{Gilmore et al.}{2007}]{gil}
G. Gilmore et al., Ap J, 663, 948 (2007).
\bibitem[\protect\citeauthoryear{Chalonge Colloquium}{2012}]{highp}
Highlights and Conclusions of the Chalonge  16th Paris Cosmology Colloquium 2012,
arXiv:1307.1847.
\bibitem[\protect\citeauthoryear{Meudon Colloquium}{2012}]{highm} 
Highlights and Conclusions of the Chalonge Meudon Workshop 2012 
arXiv:1305.7452.
\bibitem[\protect\citeauthoryear{Hogan \& Dalcanton}{2000}]{hoda}
C. J. Hogan, J. J. Dalcanton, Phys. Rev. \textbf{D62}, 063511 (2000).
\bibitem[\protect\citeauthoryear{Kormendy \& Freeman}{2004}]{kor} 
J Kormendy, K C Freeman, IAU Symposium, Sydney, 220, 377 (2004), 
arXiv:astro-ph/0407321.
\bibitem[\protect\citeauthoryear{Landau \& Lifshits}{1980}]{ll} 
L D Landau and E M Lifshits, Statistical Mechanics, Elsevier, Oxford, 1980.
\bibitem[\protect\citeauthoryear{Lapi \& Cavaliere}{2009}]{lapi}
A. Lapi, A. Cavaliere, ApJ 692, 174 (2009).
\bibitem[\protect\citeauthoryear{Lovell et al.}{2012}]{lov12}
M. R. Lovell et al., MNRAS, 420, 2318 (2012).
\bibitem[\protect\citeauthoryear{Lovell et al.}{2013}]{lovl}
M. R. Lovell et al. arXiv:1308.1399.
\bibitem[\protect\citeauthoryear{Lynden-Bell}{1967}]{lynd}
D. Lynden-Bell, Mon. Not. Roy. Astron. Soc. \textbf{136}, 101 (1967).
\bibitem[\protect\citeauthoryear{Macci\`o et al.}{2012}]{sinz}
A. Macci\`o, S. Paduroiu, D. Anderhalden, A. Schneider, B. Moore, MNRAS 424, 1105 (2012).
\bibitem[\protect\citeauthoryear{McConnachie}{2012}]{mcc}
A. W. McConnachie, AJ, 144, 4 (2012).
\bibitem[\protect\citeauthoryear{Madsen}{1990}]{mad}
J. Madsen, Phys. Rev. Lett. \textbf{64}, 2744 (1990) and Phys. Rev. \textbf{D44}, 999 (1991).
\bibitem[\protect\citeauthoryear{Martin et al.}{2008}]{martin}
N. F. Martin et al. ApJ 684, 1075 (2008).
\bibitem[\protect\citeauthoryear{Martinez et al.}{2011}]{martinez} 
G. D. Martinez et al., Ap J, 738, 55 (2011). 
\bibitem[\protect\citeauthoryear{Memola et al.}{2011}]{memo}
E. Memola et al., A\&A, 534, A50 (2011).
\bibitem[\protect\citeauthoryear{Menci et al.}{2013}]{menci}
N. Menci, F. Fiore, A. Lamastra, ApJ, 766, 110 (2013).
\bibitem[\protect\citeauthoryear {Merle}{2013}]{merle} 
A. Merle, arXiv:1302.2625, Int. J. Mod. Phys. D22, 1330020 (2013).
\bibitem[\protect\citeauthoryear{Nierenberg et al.}{2013}]{nier}
A. M. Nierenberg et al. arXiv:1302.3243, to appear in ApJ.
\bibitem[\protect\citeauthoryear{Oh et al.}{2008}]{oh}
S. H. Oh et al., AJ, 136, 2761 (2008).
\bibitem[\protect\citeauthoryear{Pacucci et al.}{2013}]{pacu}
F. Pacucci et al. arXiv:1306.0009, to appear in MNRAS Letters.
\bibitem[\protect\citeauthoryear{Papastergis et al.}{2011}]{pap}
E. Papastergis et al., Ap J, 739, 38 (2011).
\bibitem[\protect\citeauthoryear{Pe\~narrubia et al.}{2012}]{jpena}
J. Pe\~narrubia, A. Pontzen, M. Walker, S. Koposov, ApJ, 759, L42 (2012). 
\bibitem[\protect\citeauthoryear{Persic et al.}{1996}]{pers}
M. Persic, P. Salucci, F. Stel, MNRAS, 281, 27 (1996).
\bibitem[\protect\citeauthoryear{Salucci et al.}{2007}]{sal07}
P. Salucci et al. MNRAS {\bf 378}, 41 (2007).
\bibitem[\protect\citeauthoryear{Salucci}{2009}]{sal09}
P. Salucci, unpublished (2009).
\bibitem[\protect\citeauthoryear{Salucci et al.}{2012}]{sal12}
P. Salucci et al. MNRAS {\bf 420}, 2034 (2012).
\bibitem[\protect\citeauthoryear{Shao et al.}{2013}]{shao}
S. Shao et al.  MNRAS, 430, 2346 (2013).
\bibitem[\protect\citeauthoryear{Simon \& Geha}{2007}]{jdsmg} 
J. D. Simon, M. Geha,  Ap J, 670, 313 (2007) and references therein.
\bibitem[\protect\citeauthoryear{Simon et al.}{2011}]{simon11}  
J. D. Simon et al., Ap. J. 733, 46 (2011) and references therein.
\bibitem[\protect\citeauthoryear{Sommer-Larsen \& Dolgov}{2001}]{dolgov}
J. Sommer-Larsen, A. Dolgov, Ap J, 551, 608 (2001).
\bibitem[\protect\citeauthoryear{Spano et al.}{2008}]{span} 
M. Spano et al., MNRAS, 383, 297 (2008).
\bibitem[\protect\citeauthoryear{Tremaine et al.}{1986}]{thl}
S. Tremaine, M. Henon, D. Lynden-Bell, Mon. Not. Roy. Astron. Soc. \textbf{219}, 285 (1986). 
\bibitem[\protect\citeauthoryear{Tikhonov et al.}{2009}]{tikho}
A. V. Tikhonov et al., MNRAS, 399, 1611 (2009).	
\bibitem[\protect\citeauthoryear{Tonini et al.}{2006}]{toni}
Tonini, C. Lapi, A. Shankar, F. Salucci, P. 2006 ApJ, 638, 13
\bibitem[\protect\citeauthoryear{Tremaine \& Gunn}{1979}]{treg}
S. Tremaine, J. E. Gunn, Phys. Rev. Lett. 42, 407 (1979).
\bibitem[\protect\citeauthoryear{Viel et al.}{2013}]{viel}
 M. Viel et al., Phys. Rev. D88, 043502 (2013).
\bibitem[\protect\citeauthoryear{Vi\~nas et al.}{2012}]{vinas}
J. Vi\~nas, E. Salvador-Sol\'e, A. Manrique, MNRAS 424, L6 (2012).
\bibitem[\protect\citeauthoryear{Walker}{2012}]{mwal}
Matthew Walker, private communication, 2012.
\bibitem[\protect\citeauthoryear{Walker \& Pe\~narrubia}{2012}]{wp}
M. Walker, J. Pe\~narrubia, Ap. J. 742, 20 (2011).
\bibitem[\protect\citeauthoryear{Watson et al.}{2012}]{wat}
C. R. Watson et al. JCAP, 03, 018 (2012). 
\bibitem[\protect\citeauthoryear{Willman \& Strader}{2012}]{willm}
B. Willman and J. Strader, AJ, 144, 76 (2012).
\bibitem[\protect\citeauthoryear{Wolf  et al.}{2010}]{wolf10} 
J. Wolf  et al., MNRAS, 406, 1220 (2010) and references therein.
\bibitem[\protect\citeauthoryear{Woo et al.}{2008}]{woo}
J. Woo et al. MNRAS, 390, 1453 (2008).
\bibitem[\protect\citeauthoryear{Zavala et al.}{2009}]{zav}
J. Zavala et al., Ap J,	700, 1779 (2009).
\end{thebibliography}
\end{document}